\documentclass[aps,amsmath,amssymb,prd,showpacs,floatfix,preprint,superscriptaddress,nofootinbib,12pt]{article}
\usepackage{jheppub}
\usepackage{float}
\usepackage{bm}
\usepackage{mathtools}
\usepackage{tcolorbox}
\usepackage[utf8]{inputenc}
\usepackage{slashed,verbatim}
\pdfoutput=1
\usepackage{epigraph,lipsum}
\usepackage{fancyhdr}
\usepackage{commath}
\usepackage{epigraph}
\usepackage{cancel}
\usepackage{graphicx}
\usepackage{tensor}
\makeatletter
\def\@fpheader{\relax}
\makeatother

\usepackage{lipsum}

\newcommand\blfootnote[1]{%
  \begingroup
  \renewcommand\thefootnote{}\footnote{#1}%
  \addtocounter{footnote}{-1}%
  \endgroup
}

\usepackage{gensymb}

\usepackage[framemethod=tikz]{mdframed}

\usepackage{amsmath,epsfig}
\usepackage{amssymb,amsfonts}
\usepackage{latexsym}
\usepackage{graphicx}
\usepackage{commath}

\usepackage{subeqnarray}
\usepackage{xcolor}

\usepackage{graphicx}
\usepackage{ulem}

\def\be{\begin{equation}}
\def\ee{\end{equation}}
\def\bea{\begin{eqnarray}}
\def\eea{\end{eqnarray}}

\newcommand\fverb{\setbox\pippobox=\hbox\bgroup\verb}
\newcommand\fverbdo{\egroup\medskip\noindent%
                        \fbox{\unhbox\pippobox}\ }
\newcommand\fverbit{\egroup\item[\fbox{\unhbox\pippobox}]}

\newcommand{\bear}{\begin{eqnarray}}

\newcommand{\eear}{\end{eqnarray}}

\newcommand{\bsea}{\begin{subeqnarray}}
\newcommand{\esea}{\end{subeqnarray}}
\newbox\pippobox

\def\6{\partial}

\newcommand{\comments}[1]{}
\newcommand{\red}[1]{{\color{red} #1 \color{black}}}

\input Starburst.fd

\newcommand{\MB}[1]{{\red{MB: #1}}} 

\allowdisplaybreaks[3]

\newcommand{\cool}{\ensuremath{%
  \mathchoice{\includegraphics[height=2ex]{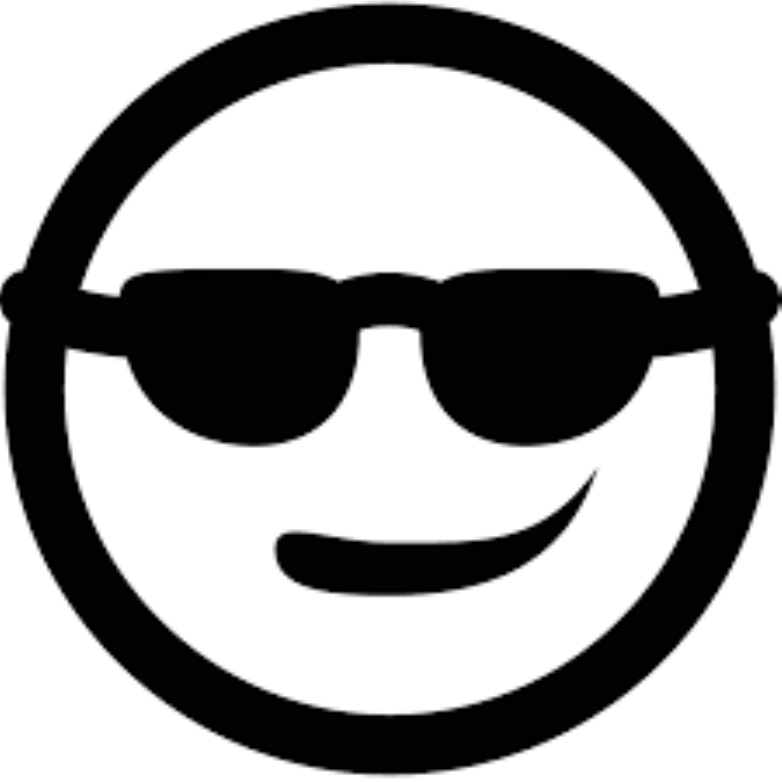}}
    {\includegraphics[height=2ex]{cool.pdf}}
    {\includegraphics[height=1.5ex]{cool.pdf}}
    {\includegraphics[height=1ex]{cool.pdf}}
}}

\newcommand{\me}{\ensuremath{%
  \mathchoice{\includegraphics[height=2.5ex]{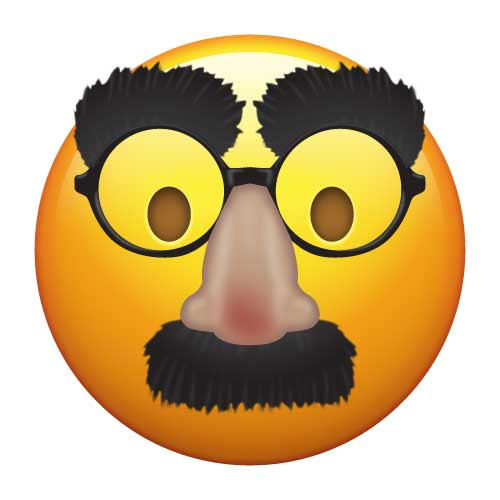}}
    {\includegraphics[height=3ex]{io.png}}
    {\includegraphics[height=2.5ex]{io.png}}
    {\includegraphics[height=2ex]{io.png}}
}}

\begin{document}

\preprint{IFT-UAM/CSIC-19-61}



\title{\centering \Huge Longitudinal Sound and  Diffusion \\ in Holographic Massive Gravity}

\author[\,\Diamond]{Martin Ammon}

\affiliation[\Diamond]{Theoretisch-Physikalisches Institut, Friedrich-Schiller-Universit\"at Jena,
Max-Wien-Platz 1, D-07743 Jena, Germany.}

\author[\,\me,\,\star]{, Matteo Baggioli}

\affiliation[\me]{Instituto de Fisica Teorica UAM/CSIC,
c/ Nicolas Cabrera 13-15, Cantoblanco, 28049 Madrid, Spain}

\author[\,\Diamond]{, Se\'an Gray}

\author[\,\Diamond\,,\,\cool]{, Sebastian Grieninger}

\affiliation[\cool]{Department of Physics, University of Washington, Seattle, WA 98195-1560, USA}

\emailAdd{martin.ammon@uni-jena.de}
\emailAdd{matteo.baggioli@uam.es}
\emailAdd{sean.gray@uni-jena.de}
\emailAdd{sebastian.grieninger@gmail.com}

\blfootnote{$\star  $ \,\,\,\url{https://members.ift.uam-csic.es/matteo.baggioli}}

\vspace{1cm}

\abstract{We consider a simple class of holographic massive gravity models for which the dual field theories break translational invariance spontaneously. We study, in detail, the longitudinal sector of the quasi-normal modes at zero charge density. We identify three hydrodynamic modes in this sector: a pair of sound modes and one diffusion mode. We numerically compute the dispersion relations of the hydrodynamic modes. The obtained speed and the attenuation of the sound modes are in agreement with the hydrodynamic predictions. On the contrary, we surprisingly find disagreement in the case of the diffusive mode; its diffusion constant extracted from the quasi-normal mode data does not agree with the expectations from hydrodynamics. We confirm our numerical results using analytic tools in the decoupling limit and we comment on some possible reasons behind the disagreement. Finally, we extend the analysis of the collective longitudinal modes beyond the hydrodynamic limit by displaying the dynamics of the higher quasi-normal modes at large frequencies and momenta.}

\maketitle

\section{Introduction}
\epigraph{$\Pi \alpha \nu \tau \alpha\,\rho \varepsilon \iota$}{Heraclitus}
\noindent Hydrodynamics and effective field theories (EFTs) are powerful tools which can be used to describe the macroscopic and low energy dynamics of physical systems. EFTs have been widely and successfully applied in a broad variety of research areas, from cosmology to particle physics, and finally to condensed matter. Symmetries are the fundamental pillars upon which such effective frameworks are built; furthermore, the symmetries of a system strongly constrain the nature of its collective excitations and their low energy dynamics. 

However, in realistic settings, and especially in the context of condensed matter systems, part of the fundamental symmetries appear to be broken. Paradigmatic examples are the spontaneous breaking of the $U(1)$ symmetry in superconducting materials, as well as the spontaneous breaking of translational (and rotational) invariance in crystals with long range order, {i.e.} systems modelled by periodic lattices. Nevertheless, even in cases where some of the symmetries are broken, effective field theory frameworks still provide an important guidance for theoretical descriptions \cite{bro}. In particular, the various symmetry breaking patterns can be used to classify the corresponding phases of matter \cite{Nicolis:2015sra}. Historically, such methods of classification have been of great importance for the understanding of the dynamics of fluids, liquid crystals, nematic crystals and ordered crystals \cite{PhysRevB.22.2514,PhysRevA.6.2401}.

Nevertheless, for systems of the types mentioned above, it is challenging to set up the hydrodynamic theory, i.e. finding a consistent and complete gradient expansion. Over the last decade, holographic dualities (also known as AdS/CFT or gauge/gravity dualities) have provided a powerful and effective tool set which can be used to tackle strongly coupled condensed matter systems  \cite{Hartnoll:2016apf,Ammon:2015wua,zaanen2015holographic}. In particular, holography has provided novel insights into hydrodynamics concerning new transport coefficients; bounds of transport coefficients; and the applicability of hydrodynamics in the presence of large gradients. Moreover, holography goes beyond hydrodynamics in the sense that it allows for direct computations of transport coefficients, which in the EFT description appear only as undetermined parameters.

Historically, holography has been focused on the description of strongly coupled fluids \cite{Policastro:2002se,Policastro:2002tn}. More recently, the attention switched to the description of strongly coupled solids or, in general a more general sense, systems with elastic properties. The long-term scope of this programme is to bring the theoretical framework closer to realistic systems in order to make concrete predictions which are testable in the lab.

The propagation of sound in compressible materials is a direct macroscopic manifestation of the spontaneous symmetry breaking (SSB) of translational invariance. Sound propagation admits the effective quasiparticle description by acoustic phonons, which are the Goldstone bosons corresponding to the spontaneous breaking of translational symmetry \cite{Leutwyler:1996er}. Sound can propagate in two different forms, transverse and longitudinal sound, which correspond to transverse and longitudinal phononic vibrational modes with dispersion relation
\begin{equation}
     \omega\,=\,\pm c_{T,L}\,k\,-\,i\,D_{T,L}\,k^2\label{disprel},
\end{equation}
where $c_{T,L}$ denotes the sound speed and $D_{T,L}$ denotes the diffusion constant. Such modes are tightly correlated to the mechanical properties of the material, in particular to its shear and bulk elastic moduli $G,K$, respectively \cite{Lubensky,landau7}.
The speed of transverse sound is determined by the simple expression 
\begin{equation}
    c_T^2\,=\,\frac{G}{\chi_{\pi\pi}},
\end{equation}
where $G$ is the shear elastic modulus and $\chi_{\pi\pi}$ the momentum susceptibility.
The property of supporting propagating shear elastic waves is usually considered the main distinction between solids and fluids.\footnote{This statement is imprecise and true only at low frequency and low momentum $\omega/T ,k/T \ll 1$. It is well known that fluids at large frequency, {i.e.} beyond the so-called Frenkel frequency, support propagating shear waves \cite{doi:10.1021/j150454a025}. Simultaneously, it is suggested by recent experiments \cite{noirez2012identification,PhysRevLett.118.215502} and theoretical developments \cite{2016RPPh...79a6502T,Baggioli:2018vfc,Baggioli:2018nnp} that the same might happen at large momentum, beyond the usually called $k$-gap.} Longitudinal sound appears both in solids and in fluids and, in two spatial directions, propagates with speed
 \begin{equation}
     c_L^2\,=\,\frac{G\,+\,K}{\chi_{\pi\pi}},
 \end{equation}
 where $K$ is the bulk elastic modulus, which is the coefficient determining the response of the system under a compressive strain.
 The diffusion constants appearing in expression \eqref{disprel} are determined by the dissipative mechanisms of the system; in a conformal system, where the bulk viscosity vanishes, they are both given in terms of the shear viscosity $\eta$. In particular, for a relativistic CFT with two spatial directions, we have
 \begin{equation}
     D_T\,=\,\frac{\eta}{\chi_{\pi\pi}}\,,\quad  D_L\,=\,\frac{1}{2}\,\frac{\eta}{\chi_{\pi\pi}}\,, \quad \mathrm{with}\quad \chi_{\pi\pi}\,=\,\varepsilon\,+\,p\,=\,s\,T,
 \end{equation}
 which can be derived using both relativistic hydrodynamics \cite{Kovtun:2012rj} and holographic techniques \cite{Policastro:2002se}. The presence of a diffusive term in the dispersion relation of the phonons \eqref{disprel} indicates that the system has also a viscous component, i.e. that it is a \textit{viscoelastic material}. In an unrealistic perfect crystal, with no dissipation nor effective viscosity, sound would propagate indefinitely, which is not realistic.
 
In a system with spontaneously broken translational symmetry the longitudinal sector contains an additional diffusive mode with dispersion relation 
\begin{equation}
     \omega\,=\,-\,i\,D_\Phi\,k^2,
\end{equation}
with diffusion constant $D_\Phi$. This diffusive mode is intimately connected with the presence of Goldstone degrees of freedom, in fact its hydrodynamic nature is simply related to the conservation of the phase of the Goldstone bosons, i.e. the phonons. The presence of such a mode can be predicted by performing an effective hydrodynamic description for crystals \cite{PhysRevA.6.2401,PhysRevB.22.2514,Delacretaz:2017zxd}. Notably, the underlying physics is very similar to that of superfluids/superconductors \cite{Davison:2016hno}, where a Goldstone diffusion mode appears as well.

All the features just mentioned can be formally derived using hydrodynamic techniques \cite{PhysRevA.6.2401,PhysRevB.22.2514,Delacretaz:2017zxd}. The analysis of the transverse modes within the holographic framework and its successful matching with the hydrodynamic predictions have already been presented in \cite{Alberte:2017oqx}. The scope of this paper is to extend the analysis to the longitudinal sector, in particular to the study of the diffusion constants of the sound and diffusive modes, in a simple holographic system with spontaneously broken translational invariance. Concretely, we use the holographic massive gravity theory introduced in \cite{Baggioli:2014roa,Alberte:2015isw} to implement the SSB breaking of translations as explained in \cite{Alberte:2017oqx}.

The hydrodynamic degrees of freedom present in the longitudinal sector have previously been investigated in a more complicated model in \cite{Andrade:2017cnc}, with more emphasis on the electric transport properties of the dual field theory. In particular the longitudinal sound mode and the diffusive mode have been identified numerically but without any systematic study or theoretical background. Related results have recently appeared in  \cite{Baggioli:2019aqf} in the presence of electromagnetic interactions and in \cite{Baggioli:2019abx} for holographic fluid models. Other recent works have analysed the origin of this extra diffusive mode in similar holographic models with a global $U(1)$ symmetry \cite{Donos:2019hpp,Donos:2019txg}. In this work, thanks to the simplicity of the model at hand, we improve the analysis. In particular:
\begin{enumerate}
    \item We study in the dynamics of the longitudinal collective modes dialing the SSB scale.
    \item We compare numerical results obtained from the holographic model with the hydrodynamics expectations.
    \item We perform analytic computations in the decoupling limit, where the Goldsone mode's dynamics are independent of the gravitational dynamics.
    \item We numerically describe the higher modes, extending the analysis beyond the hydrodynamic limit towards large frequencies and momenta, $\omega/T,k/T \gg 1$.
\end{enumerate}

From the numerical study we obtain the expected hydrodynamic modes, namely the longitudinal sound and the diffusive mode. Both modes display the correct dispersion relation. Moreover, the propagation speed and the attenuation constant of the sound mode are in agreement with the hydrodynamic predictions \cite{PhysRevA.6.2401,PhysRevB.22.2514,Delacretaz:2017zxd} at every value of the SSB scale. Instead, we find a disagreement between the diffusion constant of the diffusive mode and the value predicted by hydrodynamics. We will comment on this puzzle below. Our results indicate that a complete understanding of these holographic models, in terms of an hydrodynamic effective description, may still be lacking.

\section{Holographic Set-up}\label{sec:setup}
We consider generic \textit{viscoelastic} holographic massive gravity models \cite{Baggioli:2014roa,Alberte:2015isw} defined by the following four-dimensional gravitational action in the bulk
\begin{equation}\label{action}
S\,=\, M_P^2\int d^4x \sqrt{-g}
\left[\frac{R}2+\frac{3}{\ell^2}-  m^2 \, V(X)\right] \, ,
\end{equation}
with $X \equiv \frac{1}{2} \, g^{\mu\nu} \,\partial_\mu \phi^I \partial_\nu \phi^I$ and $m$ is a coupling which relates to the graviton mass \cite{Alberte:2015isw}. Here, $\phi^I$ are St\"uckelberg scalars which admit  a radially constant profile, $\phi^I=x^I$ with $I=1,2$. In the dual field theory the St\"uckelberg fields represent scalar operators breaking the translational invariance of the system. Whether the translational symmetry breaking is explicit or spontaneous depends crucially on the bulk potential of the scalar fields. In particular, depending on the choice of the potential $V(X)$, the bulk solution $\phi^I=x^I$ may either correspond to the non-normalisable mode and hence it acts as a source term for the dual operators  $\Phi^I$; or it corresponds to the normalisable mode and hence it gives rise to a non-zero expectation value $\langle \Phi^I \rangle$. In the following we will concentrate on potentials of the type
\begin{equation}
    V(X)\,=\,X^N\,,\quad \textrm{with} \quad N\,>\,5/2\,,
\end{equation}
which in a simple way realise the spontaneous breaking of translational invariance \cite{Alberte:2017oqx}. 

The thermal state is dual to an asymptotically AdS black hole geometry which in Eddington-Finkelstein (EF) coordinates is described by the metric
\begin{equation}
\label{backg}
ds^2=\frac{1}{u^2} \left[-f(u)\,dt^2-2\,dt\,du + dx^2+dy^2\right] \, ,
\end{equation}
where $u\in [0,u_h]$ is the radial holographic coordinate. The conformal boundary is located at $u=0$ and the horizon $u_h$ is defined by $f(u_h)=0$. Hence the emblackening factor $f(u)$  takes the simple form
\begin{equation}\label{backf}
f(u)= u^3 \int_u^{u_h} dv\;\left[ \frac{3}{v^4} -\frac{m^2}{v^4}\, 
V(v^2) \right] \, .
\end{equation}
The corresponding temperature of the dual QFT reads
\begin{equation}
T=-\frac{f'(u_h)}{4\pi}=\frac{6 -  2 m^2 V\left(u_h^2 \right) }{8 \pi u_h} \, , \label{eq:temperature}
\end{equation}
while the entropy density is $s=2\pi/u_h^2$. For simplicity, and without loss of generality, we will fix $u_h=1$ in the rest of the manuscript.

The heat capacity is $c_V=\partial \varepsilon / \partial T$, with $\varepsilon$ being the energy density and $T$ being the temperature, and it has been studied in \cite{Baggioli:2015gsa,Baggioli:2018vfc}. The dual field theory has been shown to possess viscoelastic features and was studied in detail in \cite{Baggioli:2015zoa,Baggioli:2015gsa,Baggioli:2015dwa,Alberte:2016xja,Alberte:2017cch,Alberte:2017oqx,Ammon:2019wci}.\footnote{The viscoelastic nature of these holographic systems and their features have recently inspired important results in the understanding of amorphous solids and glasses \cite{Baggioli:2018qwu}.}

It is important to stress that our model only has one dimensionless scale controlling the strength of the spontaneous translational symmetry breaking. More specifically, we have
\begin{equation}
    \langle SSB \rangle \sim\,m/T \, ,
\end{equation}
where the limiting case $m=0$ corresponds to the translationally invariant system. 

In the following we consider fluctuations around the thermal equilibrium state and  compute the quasi-normal modes. The transverse sector of the fluctuations has been studied in detail in \cite{Alberte:2017oqx} and it revealed the presence of damped transverse phonons whose speed are in perfect agreement with the theoretical predictions from hydrodynamics. Within this manuscript, we continue this analysis by considering instead the quasi-normal modes within the longitudinal sector.

\section{Hydrodynamics and Longitudinal Quasi-Normal Modes}\label{sec:hydro}
In this section we will study the hydrodynamic modes appearing in the longitudinal sector of the quasi-normal mode spectrum.\footnote{By ``hydrodynamic modes'' we mean the modes present in the system within the range $\omega/T$, $k/T \ll 1$.}  Furthermore, we compare the dispersion relations of the quasi-normal modes to the corresponding, analytic, hydrodynamic expectations.

We defer the relevant equations of motion for the fluctuations and the technicality concerning the holographic computation to appendices \ref{app1} and \ref{app2}. Moreover, in appendix  \ref{hydroapp} we review and (partially) extend the hydrodynamic effective description of systems with broken translations following the lines of \cite{Delacretaz:2017zxd}.\footnote{All the physics of this section is already included in \cite{Delacretaz:2017zxd}; nevertheless we will display results beyond the specific approximations assumed in \cite{Delacretaz:2017zxd}.}

\begin{figure}
    \centering
    \includegraphics[width=9cm]{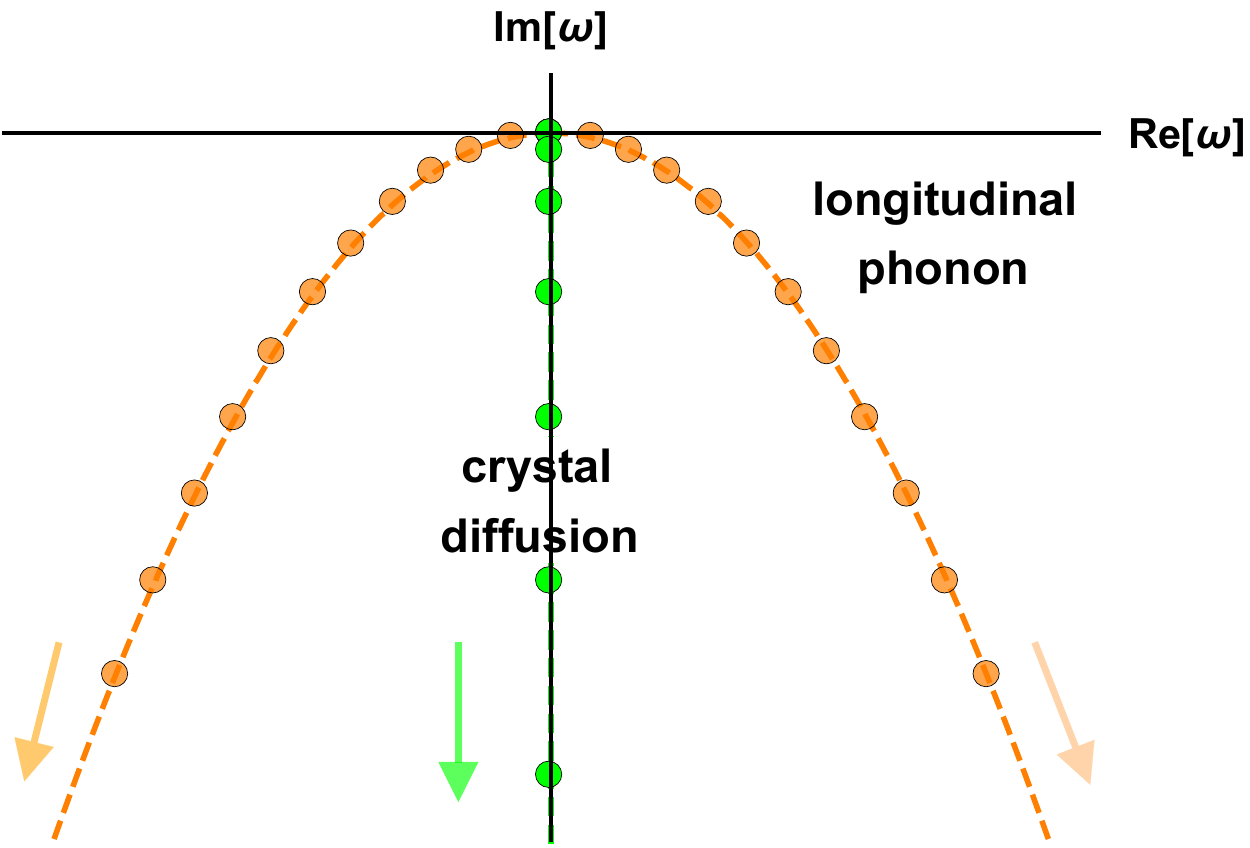}
    \caption{A sketch of the two hydrodynamic modes appearing in the longitudinal sector of the QNMs of our system. In orange the longitudinal damped sound $\omega\,=\,\pm c_L k-i\, D_p\, k^2$ and in green the diffusive mode $\omega=-i\,D_\Phi\, k^2$. The arrows indicate the tendency when increasing the momentum $k$.}
    \label{fig:cartoon}
\end{figure}

\subsection{Hydrodynamic Regime of Quasi-Normal Modes}
First we analyse the quasi-normal modes in the low-frequency and long-wavelength regime. In particular the spectrum exhibits:
\begin{enumerate}
    \item A pair of longitudinal damped sound modes with dispersion relation
    \begin{equation}\label{mode1}
        \omega\,=\,\pm c_L\,k\,-i\,D_p\,k^2\,+\,\dots \, ,
    \end{equation}
    with the speed $c_L$ and the diffusion constant $D_p$.\footnote{Sometimes people refer to the sound attenuation constant $\Gamma_S$ which corresponds to $\Gamma_S \equiv 2\,D_p$.} The ellipsis stands for higher momenta corrections. The mode \eqref{mode1} is exactly a longitudinal damped phonon mode, which is expected both in fluids and solids.
    \item A diffusive mode with dispersion relation
    \begin{equation}
        \omega\,=\,-\,i\,D_\Phi\,k^2\,+\,\dots \, ,
    \end{equation}
    where $D_\Phi$ is the diffusion constant. The ellipsis stands for higher momenta corrections. The presence of such a diffusive mode is typical of systems which break translational invariance spontaneously. This hydrodynamic mode can be viewed as the diffusive mode of the Goldstone parallel to the momentum.
\end{enumerate}
\begin{figure}
\centering
\includegraphics[width=7.5cm]{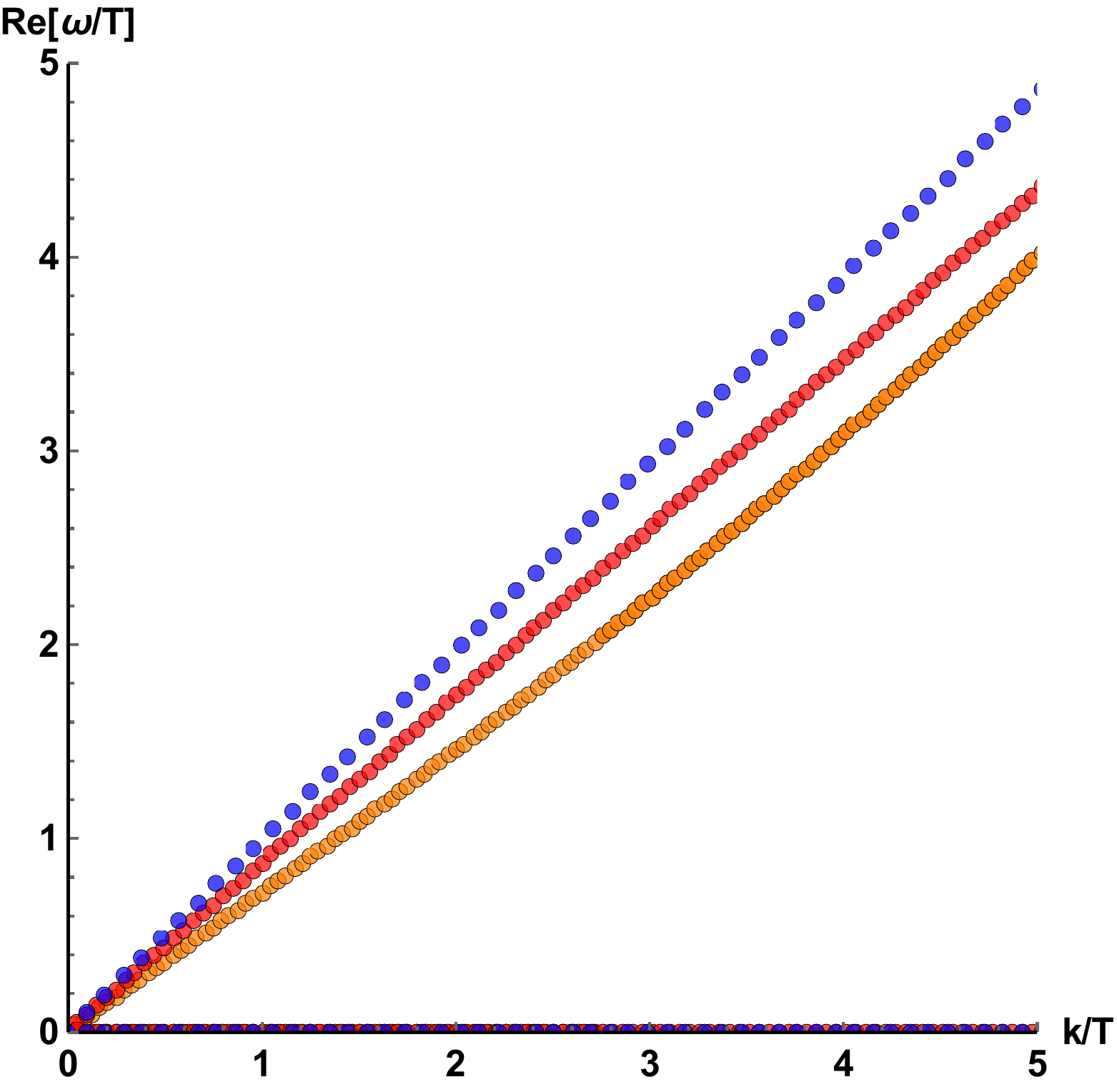}
\quad
\includegraphics[width=7.5cm]{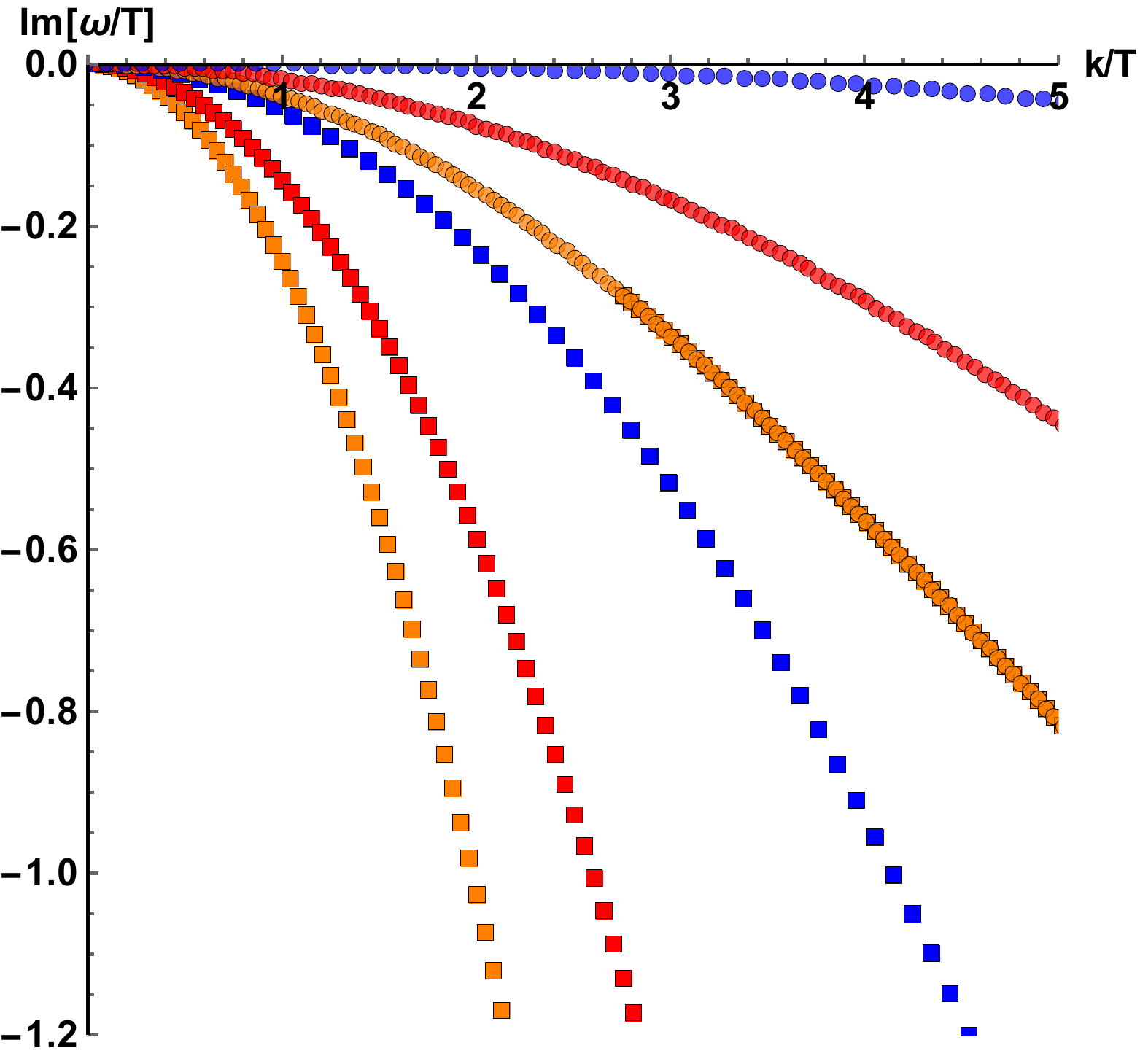}
\caption{A snapshot of the typical dispersion relations for the sound mode (disks) and the diffusive mode (squares). The specific example pertains to the potential $V(X)=X^3$ for $m/T=0,3,11$ (from orange to blue).}
\label{examplefig}
\end{figure}
The two modes presented above are shown in figure \ref{examplefig}, for the specific potential $N=3$ and for different values of the dimensionless parameter $m/T$. We obtain similar results for other values of $N$ with $N>5/2$ (see figure \ref{fig:cartoon} for a schematic representation of the hydrodynamic modes of our system).

\subsection{Correlators of the Goldstone Modes}
Before comparing the numerically obtained dispersion relations of the low-lying quasi-normal modes to the expected dispersion relations of hydrodynamics, let us discuss the retarded Green's functions which will be relevant for the future analysis. We are particularly interested in correlators appearing due to the SSB of translations, i.e. those containing the Goldstone fields $\Phi^I$ which are dual to the St\"uckelberg fields $\phi^I$ in the bulk. To be more precise, we discuss the retarded Green's functions $ \mathcal{G}_{\pi^I \Phi^J}(\omega, \vec{k})$ and $\mathcal{G}_{\Phi^I \Phi^J}(\omega, \vec{k})$, 
where $\Phi^I$ and $\pi^I$ (with $I=1,2$) are the two Goldstone modes and the momentum operators, respectively. 

\begin{figure}
    \centering
    \includegraphics[width=4.7cm]{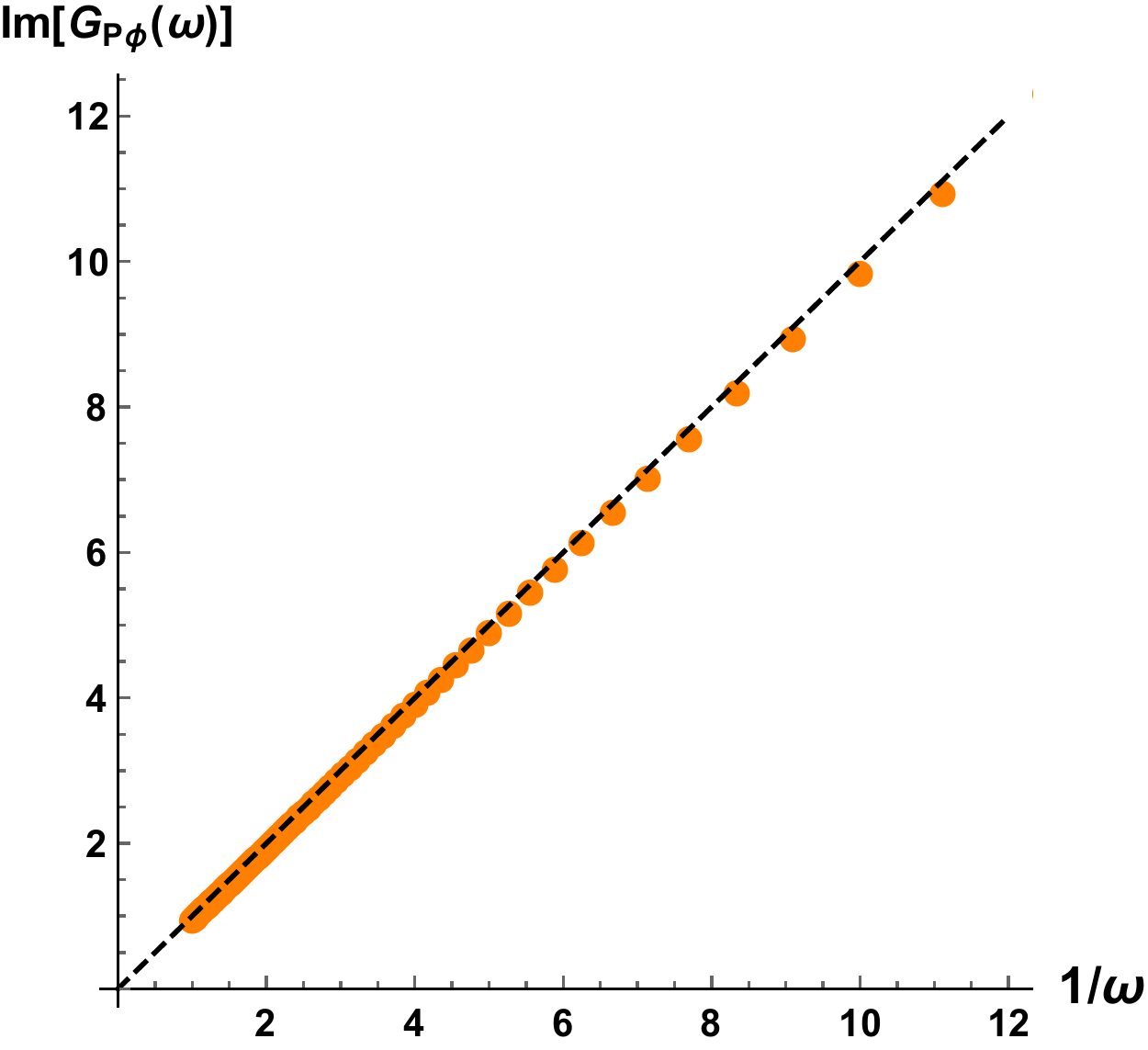}\quad
    \includegraphics[width=4.7cm]{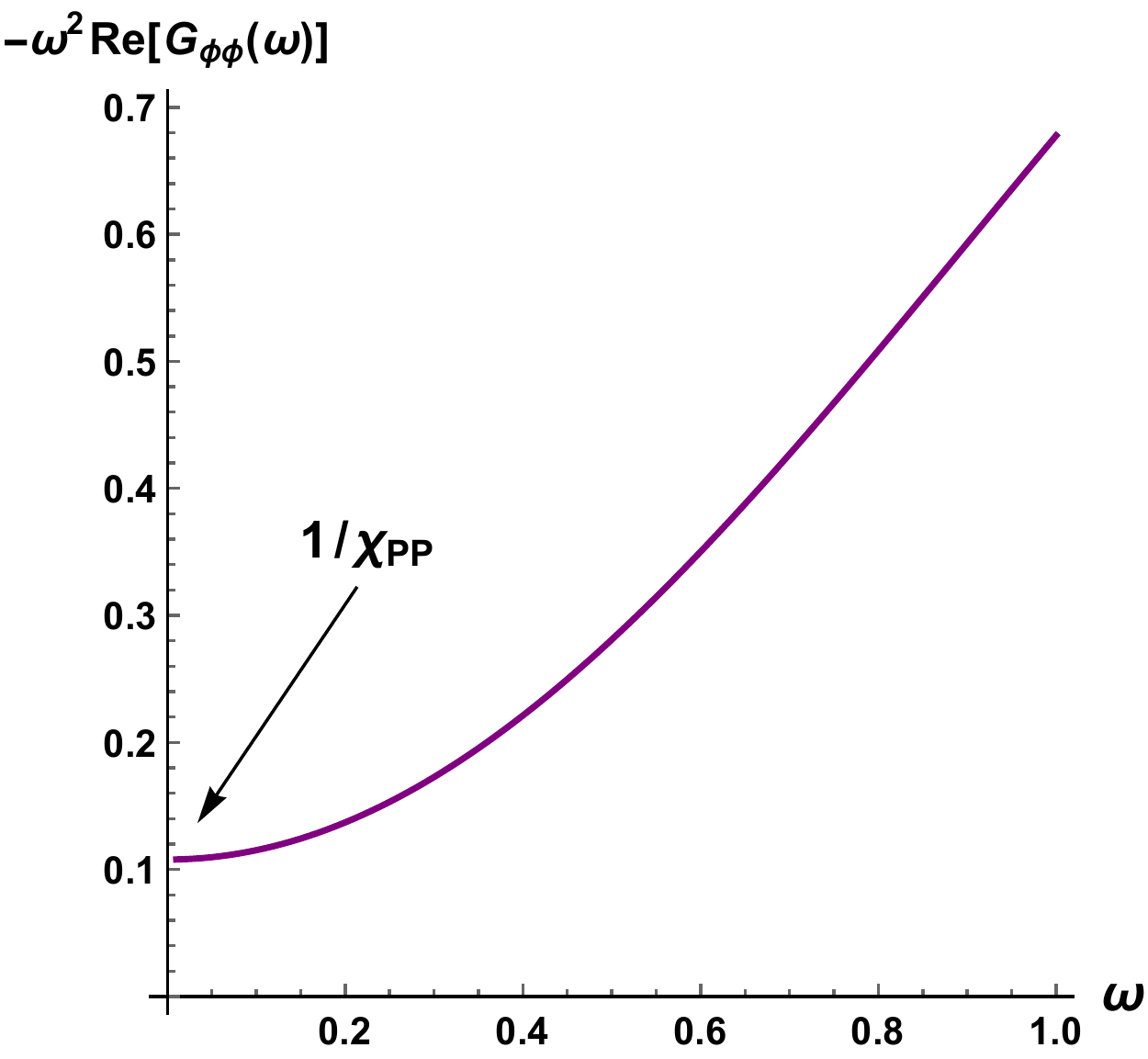}\quad
    \includegraphics[width=4.7cm]{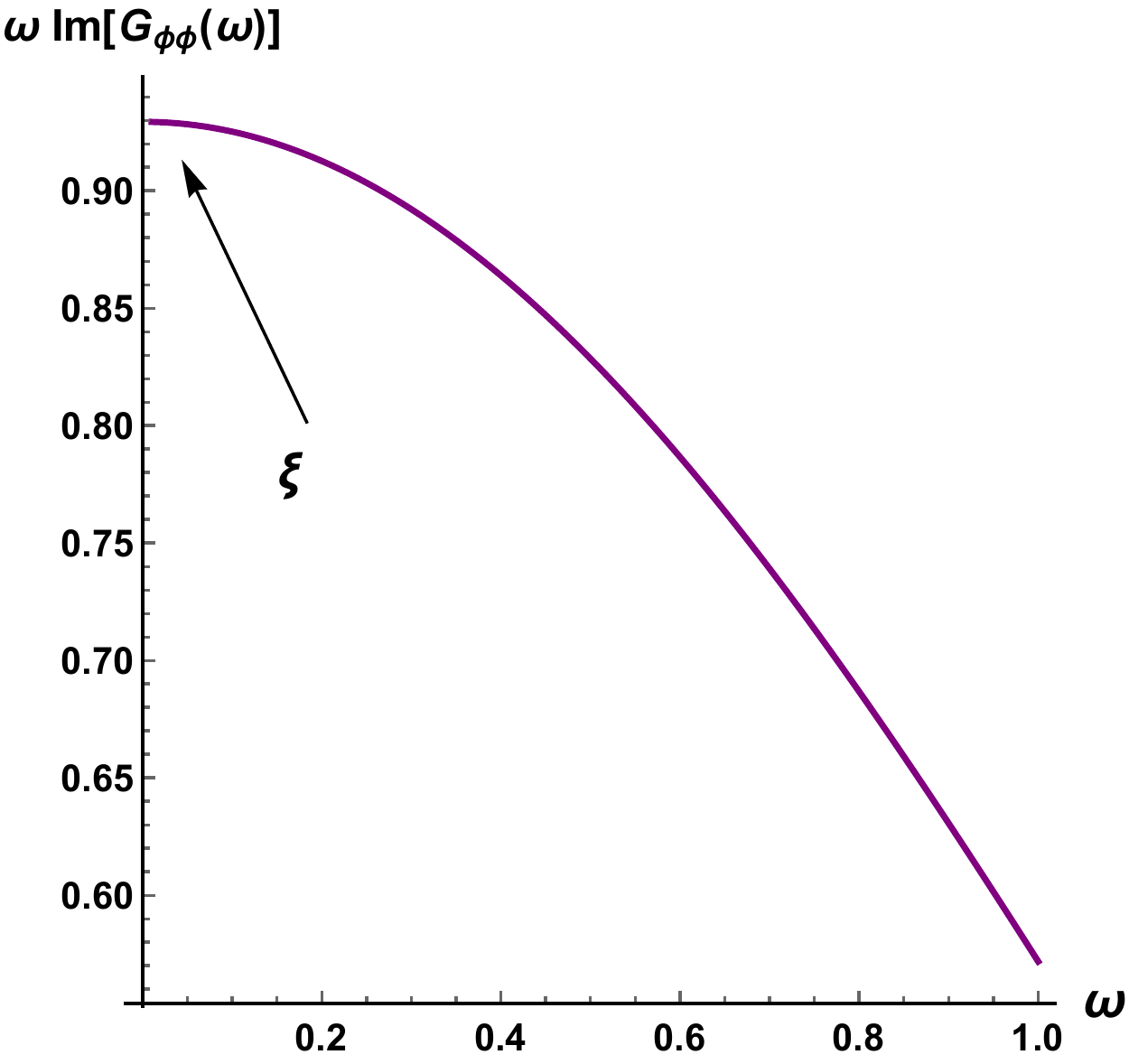}
    \caption{The correlator of the Goldstone operator $\Phi$ as a function of the frequency for $m/T=1$ and $N=3$. \textbf{Left: } The imaginary part of the mixed correlator $\langle P \Phi \rangle$. The dashed line guides the eye towards the $\sim 1/\omega$ behaviour. \textbf{Center: }The Real part of the $\langle \Phi \Phi \rangle$ correlator multiplied by $\omega^2$. The zero frequency value coincides with the inverse of the momentum susceptibility. \textbf{Right: }The imaginary part of the $\langle \Phi \Phi \rangle$ correlator multiplied by the frequency $\omega$. The zero frequency value coincides with the parameter $\xi$.}
    \label{fig:checkcorrelators}
\end{figure}

Using the hydrodynamic description of phases with spontaneously broken translational symmetry, found in \cite{PhysRevB.22.2514, Delacretaz:2017zxd} and repeated in our Appendix \ref{hydroapp}, we can derive the form of the previous Green's functions at low energy/frequency
\begin{equation}
    \mathcal{G}_{\pi^I  \Phi^J}(\omega,k)\,=\,\frac{i}{\omega} \, \delta^{IJ}\,+\,\mathcal{O}(k^2) \,,\quad \mathcal{G}_{\Phi^I \Phi^J}(\omega,k)\,=\left(-\frac{1}{\omega^2\,\chi_{\pi\pi}}\,+\,\xi\,\frac{i}{\omega}\right) \, \delta^{IJ}\,\,+\,\mathcal{O}(k^2) \label{corrhydro}
\end{equation}
where $\chi_{\pi\pi}$ is the momentum susceptibility and $\xi$ relates to the  Goldstone diffusion.

The holographic correlators can be derived from the equations for the fluctuations of the bulk fields using the standard holographic dictionary \cite{Skenderis:2002wp}; see Appendix \ref{app2} for further details, in particular regarding the mapping between the scalar St\"uckelberg fields $\phi^I$ and the Goldstone operators $\Phi^I$. 

\begin{figure}
    \centering
    \includegraphics[width=0.6\linewidth]{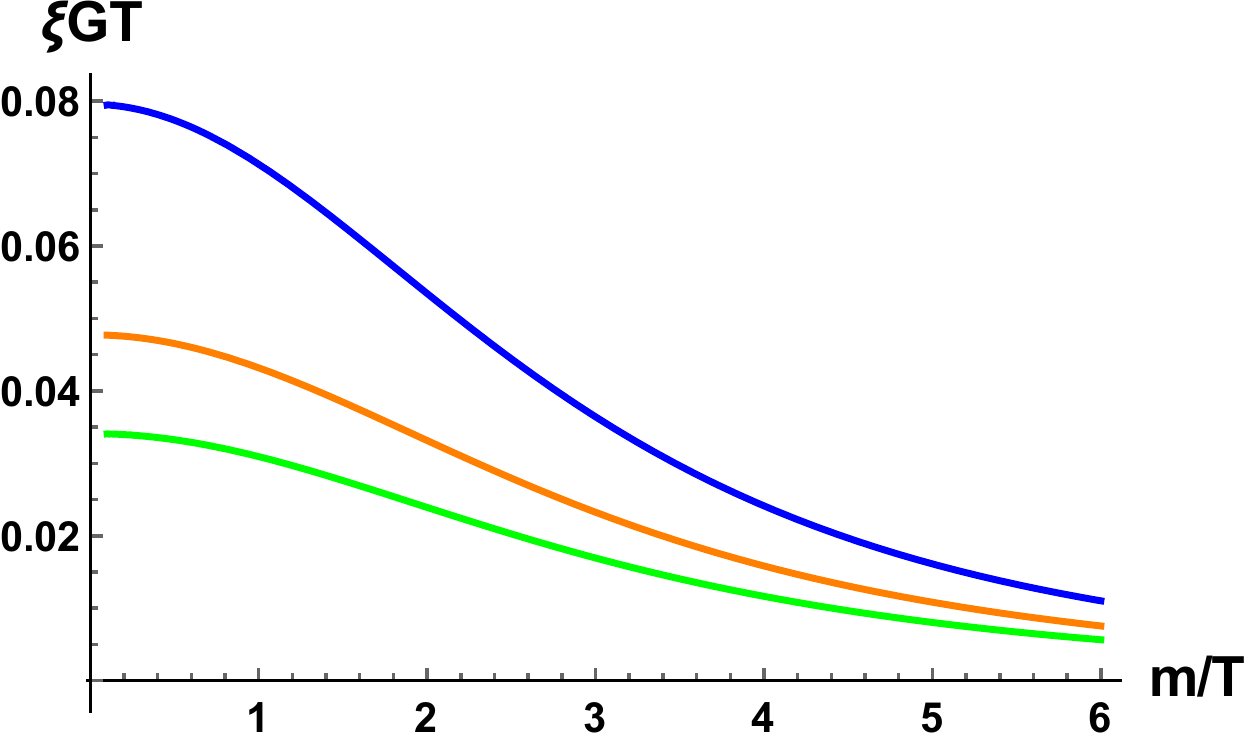}
    \caption{The value of the $\xi$ parameter \eqref{xidef}, opportunely made dimensionless using the shear modulus $G$ and the temperature $T$, as a function of the SSB parameter $m/T$ for $N=3,4,5$ (from blue, top; to green, bottom).}
    \label{figxi}
\end{figure}

We have conducted numerical studies of the behaviour of the Green's functions \eqref{corrhydro} within the context of our holographic model and for various potentials $V(X)=X^N$ with $N>3$. One example of the results is shown in figure \ref{fig:checkcorrelators}. We find perfect agreement between the Green's functions obtained from holography and from hydrodynamics. The transport coefficient $\xi$ can be read off from the imaginary part of the Green's function $\mathcal{G}_{\Phi^I \Phi^J}$ as 
\begin{equation}
    \xi\,=\,\lim_{\omega \rightarrow 0} \omega \,\lim_{k\rightarrow 0}
    \,\mathrm{Im}\left[\mathcal{G}_{\Phi \Phi}(\omega, {k})\right]\,\label{xidef},
\end{equation}
where, in this limit,  it is no longer necessary to to distinguish between $\Phi^{(1)}$ or $\Phi^{(2)}$.\footnote{At zero momentum, $k=0$, and for isotropic backgrounds, there is no distinction between the various directions.}
A plot of the $\xi$ parameter for various $N$ is shown in figure \ref{figxi}. Furthermore, the numerical results for $\xi$ are in perfect agreement with the horizon formula
\begin{equation}
    \xi\,=\,\frac{4\,\pi\,s\,T^2}{N\,2\,m^2\,\chi_{\pi\pi}^2},
\end{equation}
given in \cite{Amoretti:2018tzw} and derived in \cite{Amoretti:2019cef}, which here is used at zero chemical potential $\mu=0$.\footnote{Notice the different notations.  The map between the conventions is
\begin{equation}
    \left(\frac{\xi}{G}\right)_{their}\,\equiv \xi\,,\quad k^2 \rightarrow 2\,m^2\,,\quad Y_h\,\rightarrow\,N\,.
\end{equation}}

\subsection{Matching to Hydrodynamics}
 Let us compare the dispersion relations obtained from the quasi-normal mode analysis presented above with the theoretical expectations from hydrodynamics. For details regarding the hydrodynamic analysis see Appendix \ref{hydroapp}.

\subsubsection{Sound Mode}\label{sec:soundmode}
First we analyse the dispersion relation of the propagating longitudinal mode \eqref{mode1}, starting with its speed $c_L$. The hydrodynamic analysis performed in Appendix \ref{hydroapp} predicts the speed $c_L$ to be given by
\begin{equation}\label{eq:hydrosound}
    c_L^2\,=\,\frac{\partial p}{\partial \varepsilon}\,+\,\frac{\kappa\,+\,G}{\chi_{\pi\pi}},
\end{equation}
where $\varepsilon$ is the energy density; $p$ is the thermodynamic pressure; and $\chi_{\pi\pi}$ is the momentum susceptibility. Moreover, $\kappa$ and $G$ are the bulk and shear elastic modulus, respectively. Both $\kappa$ and $G$ vanish in the absence of translational symmetry breaking and the formula \eqref{eq:hydrosound} reduces to the well-known speed formula for sound $c_L^2 = \partial p/\partial\varepsilon$; or $c_L^2\,=1/d$ in the special case of a $d$-dimensional CFT. 

Reminiscent of solids, we can re-express the speed of the longitudinal sound modes as 
\begin{equation} \label{eq:hydrosound2}
    c^2_{L} = \frac{G+K}{\chi_{\pi\pi}} \, , 
\end{equation}
where $K$ is the thermodynamic bulk modulus obtained from the inverse of the compressibility
\begin{equation}
    K\,\equiv\,-\,\mathcal{V}\,\frac{\partial \mathcal{P}}{\partial \mathcal{V}}\label{totalK} \, ,
\end{equation}
where $\mathcal{V}$ is the volume of the system and $\mathcal{P}$ is the mechanical pressure defined by $\mathcal{P}\equiv T_{ii}$ (no sum over $i$ implied). Let us notice that, in our model
\begin{equation}
    \mathcal{P}\,=\,p\,+\,2\,\kappa,
\end{equation}
which means that the thermodynamic and mechanical pressures coincide only at zero SSB, $m/T=0$. In order to establish an equivalence between \eqref{eq:hydrosound} and \eqref{eq:hydrosound2} we must demand
\begin{equation}
K\,=\,\frac{\partial p}{\partial \varepsilon}\,\chi_{\pi\pi}\,+\,\kappa \, .\label{concon}
\end{equation}
In other words,  the first term in the above formula is a contribution which is finite even in the absence of SSB, and it is indeed what determines the speed of sound in a pure relativistic CFT. The second term captures additional effects due to spontaneous symmetry breaking.  

Let us now determine the coefficients $G$, $K$ and $\kappa$ within our model. The elastic shear modulus is given in terms of the Kubo formula\footnote{For our definition of the retarded Green's function see appendix \ref{app2}.}
\begin{equation}\label{defd}
    G\,=\,\lim_{\omega \rightarrow 0}\,\lim_{k \rightarrow 0} \, \textrm{Re}  \left[\mathcal{G}_{T_{xy}T_{xy}}(\omega,k)\right],
\end{equation}
and has been discussed in \cite{Alberte:2015isw,Alberte:2016xja,Alberte:2017cch,Alberte:2017oqx,Baggioli:2018bfa} within the context of the holographic model considered here. 

The bulk modulus $\kappa$ can be determined in various ways. Using the underlying conformal invariance of the dual QFT, as well as the structure of the energy-momentum tensor, the bulk modulus $\kappa$ reads \cite{Baggioli:2018bfa} 
\begin{equation}
    \kappa\,=\,\frac{3\,\varepsilon\,-\,2\,s\,T}{4}.
\end{equation}
In fact, using $\chi_{\pi\pi} = 3\varepsilon/2 $ as well as\footnote{See also \cite{oriol} for more discussions on this point.}  \begin{equation}
    K\,=\,\frac{3}{4}\,\varepsilon,
\end{equation}
which is valid for two-dimensional conformal solids, we can check that the consistency relation \eqref{concon} is indeed satisfied. Finally, we may also compute $\kappa$ by a Kubo formula; for more details see appendix \ref{sec:Kuboformulae}.

\begin{figure}
\centering
\includegraphics[width=7.5cm]{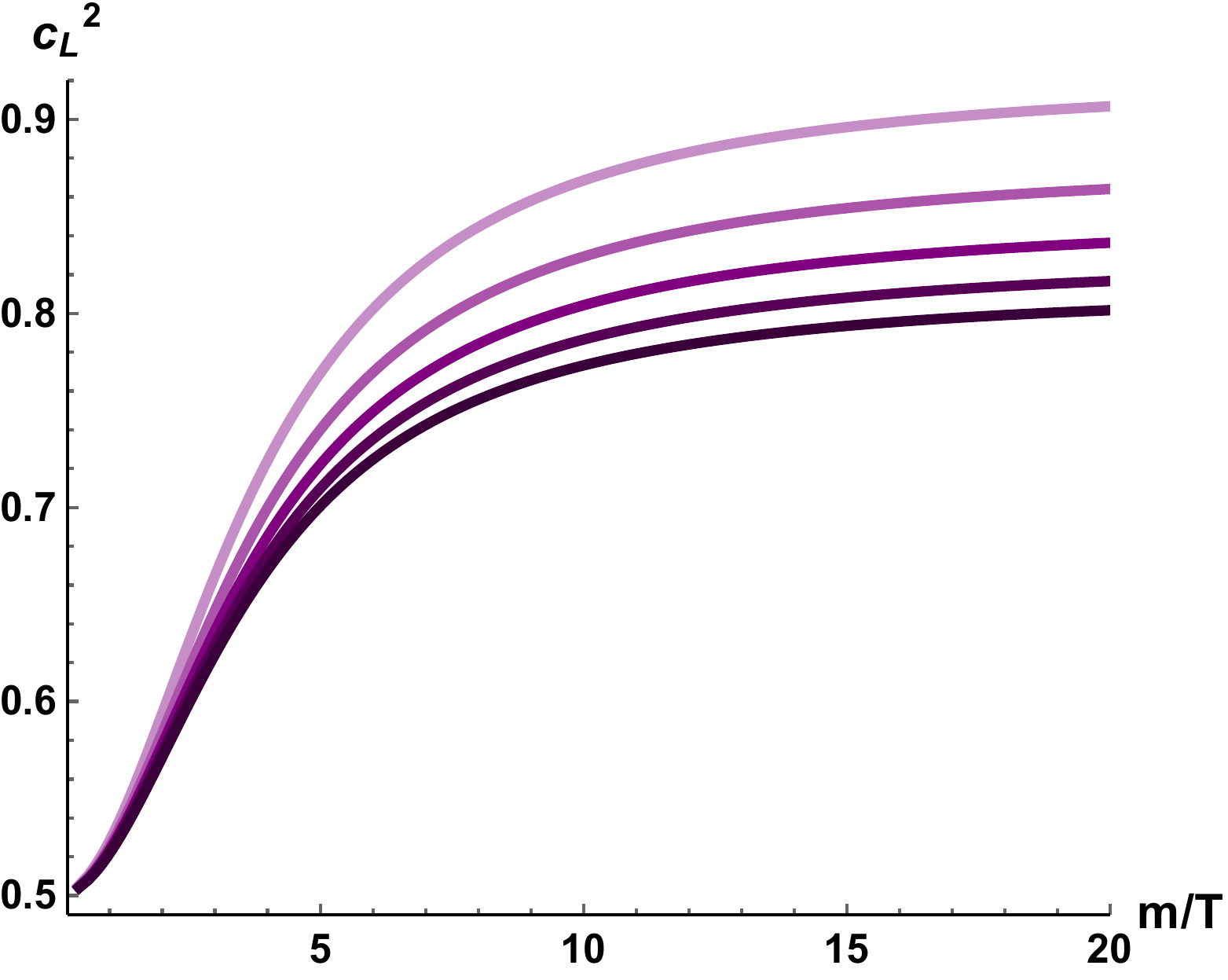}
\quad
\includegraphics[width=7.5cm]{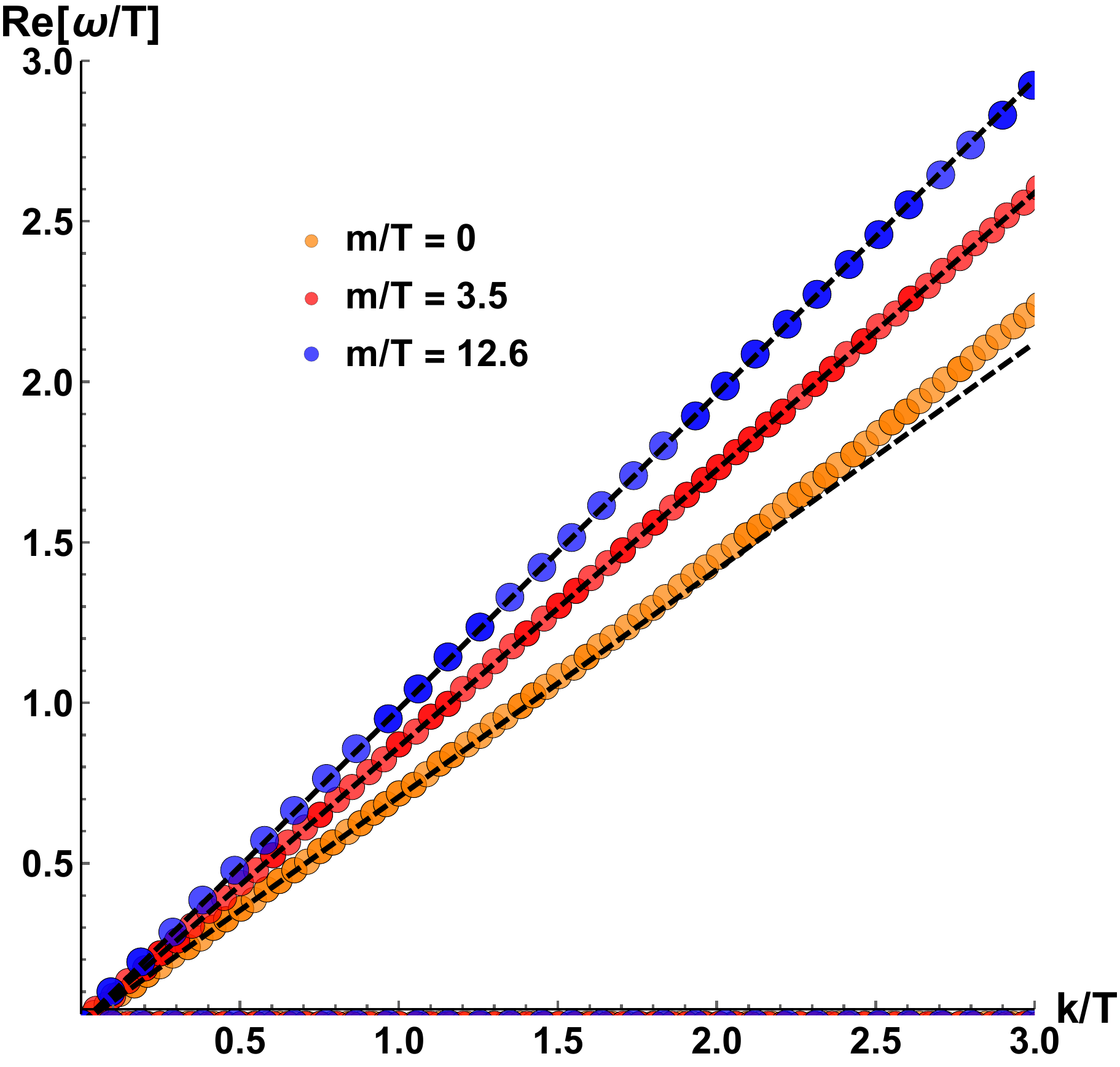}
\caption{\textbf{Left: }Longitudinal phonons speed for various potentials $N=3,4,5,6,7,8$ (from lighter to darker color) in function of the dimensionless parameter $m/T$. \textbf{Right: }Real part of the dispersion relation of the longitudinal phonons for the potential $N=3$ and various $m/T\in \{0, 3.5, 12.6\}$. The dashed black lines are the comparisons with the theoretical formula \eqref{eq:hydrosound}; the agreement is perfect.}
\label{speedfig}
\end{figure}

Collecting all the above formulae we end up re-discovering a relation between the transverse and longitudinal speeds of sound put forward in \cite{Esposito:2017qpj,Alberte:2018doe}, i.e. that we have 
\begin{equation}
    c_L^2\,=\,c_T^2\,+\,\frac{1}{2}
\end{equation}
in a two-dimensional conformal solid, for any value of the SSB parameter $m/T$. 

The results for the sound speed are shown in figure \ref{speedfig}. In the left panel we show the value of $c_L^2$ from equation \eqref{eq:hydrosound} as a function of the dimensionless symmetry breaking scale $m/T$, for various potentials of the form $V(X)=X^N$. At $m/T=0$ we recover the CFT result $c_L^2=1/2$, which acts as a lower bound for the speed of longitudinal sound. Increasing $m/T$, the speed $c_L$ grows until it reaches a constant value at large $m/T \rightarrow \infty$. Depending on the choice of the potential, the speed of sound might become superluminal and the system can exhibit causality pathologies (see \cite{oriol}). However, this is not the case for the choice $V(X)=X^N$, provided $N\geq3$. In the right panel we compare the numerical data extracted from the QNMs and the theoretical formula written down in equation \eqref{eq:hydrosound}. Within the hydrodynamic limit $k/T \ll 1$ the agreement is excellent. 
\begin{figure}[h]
\centering
\includegraphics[width=7.5cm]{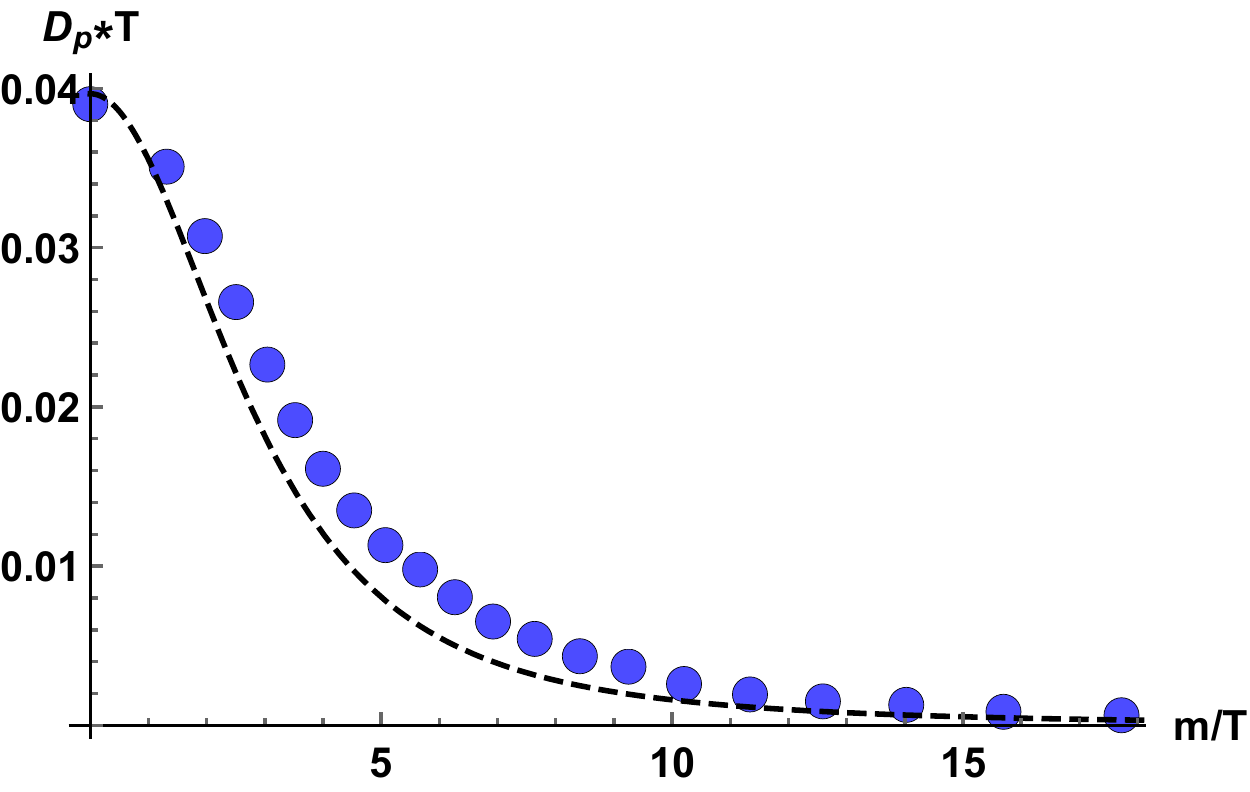}
\quad
\includegraphics[width=7.5cm]{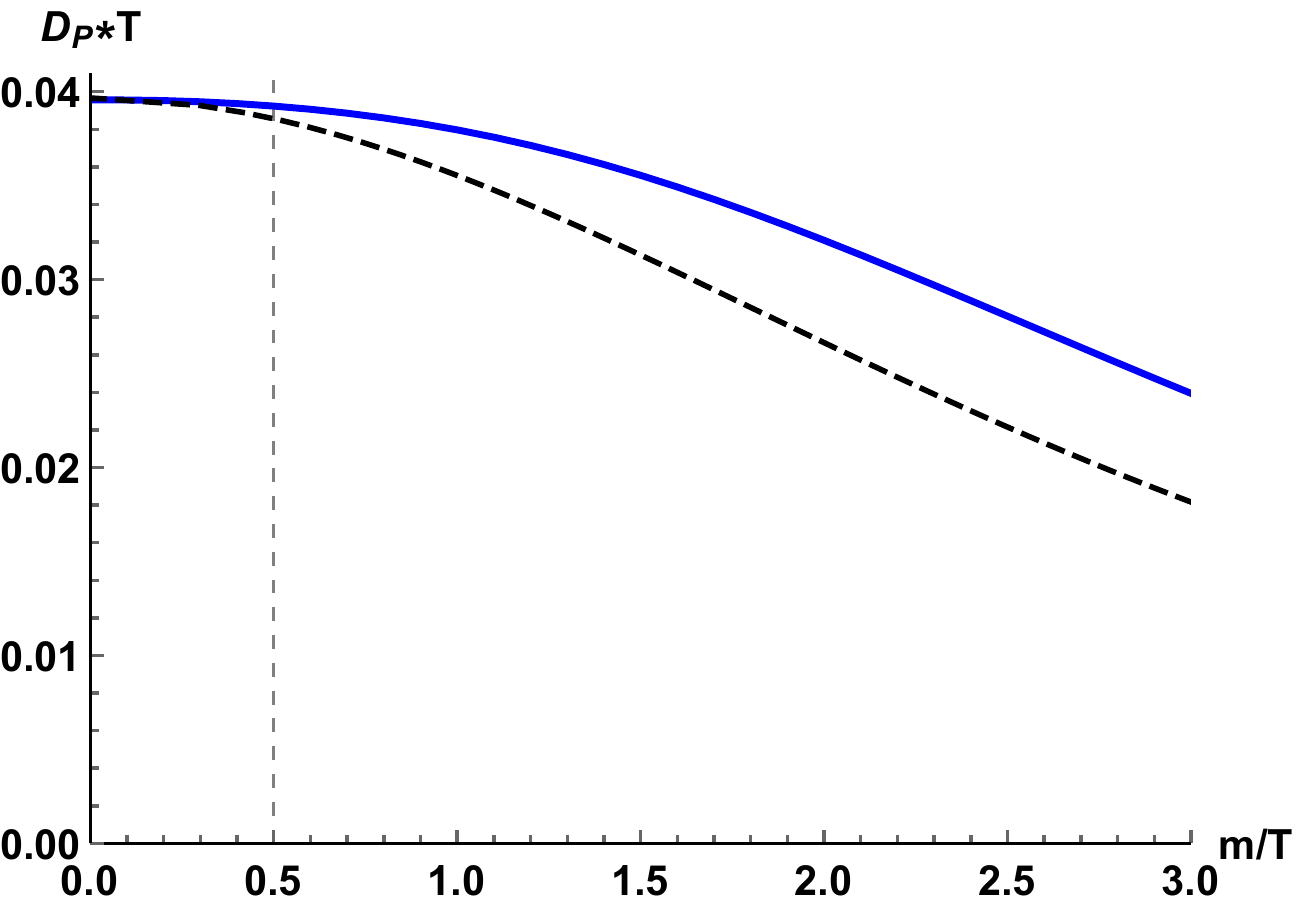}
\caption{\textbf{Left: } The longitudinal phonons' diffusion constant $D_p$ as a function of the dimensionless parameter $m/T$ for $V(X)=X^3$. \textbf{Right: } The comparison between the numerical data (solid line) and the theoretical expectations for $m/T \ll 1$.}
\label{phondifffig}
\end{figure}

Let us now move on and discuss the dissipative part of the phonon dispersion relation. Because of finite temperature effects, or in other words due to the presence of an event horizon and its associated shear viscosity $\eta$, the longitudinal phonons obtain a finite diffusive damping $D_p$. The same phenomenon occurs also in the transverse sector, the effects of which were studied in \cite{Alberte:2017oqx}. The interplay of elasticity (a propagating term) and viscosity (damping) qualify the system at hand to represent the gravity dual of a viscoelastic material. 
Hydrodynamics provides us with the formula for the diffusion constant, which reads
\begin{equation}
    D_p = \frac12 \frac{\eta}{\chi_{\pi\pi}} + \frac12 \frac{c_V (\kappa+G)^2\xi - (\kappa+G) c_V (\partial p / \partial \varepsilon) T \gamma_2 + (\partial p / \partial \varepsilon) \bar{\kappa}_0 \chi_{\pi\pi} -\gamma_2 (\kappa+G) \chi_{\pi\pi}}{c_V (\kappa + G + (\partial p / \partial \varepsilon) \chi_{\pi\pi})},
\end{equation}
where we define the specific heat $c_V\equiv \partial \varepsilon/\partial T$ and the parameters $\eta$, $\xi$, $\bar{\kappa}_0$ and $\gamma_2$  as 
\begin{align}
\eta\,&=-\,\lim_{\omega\, \rightarrow 0}\,\frac{1}{\omega}\,\lim_{k \to 0}\,\mathrm{Im}\left[\mathcal{G}^R_{T_{xy}T_{xy}}\left(\omega,k\right)\,\right],\\
 \xi\,&=\,\lim_{\omega\, \rightarrow 0}\,\omega\,\lim_{k \to 0}\,\mathrm{Im}\left[\mathcal{G}^R_{\Phi\Phi}\left(\omega,k\right)\,\right],\\
    \bar{\kappa}_0 &= -\lim_{\omega \to 0} \omega \lim_{k \to 0} \frac{1}{k^2} \mathrm{Im}\left[\mathcal{G}^R_{\varepsilon\varepsilon}(\omega,k)\right], \label{eq:kuboKappaMain}\\
    T \gamma_2 &= -\lim_{\omega \to 0} \omega \lim_{k \to 0} \frac{1}{k} \mathrm{Re}\left[\mathcal{G}^R_{\varepsilon\Phi}(\omega,k)\right], \label{eq:kuboGammaMain}
\end{align}
where $\Phi$ is the Goldstone operator (parallel to $\vec{k}$) of the dual field theory (not to be confused with the scalar bulk field $\phi$). 

\begin{figure}
    \centering
    \includegraphics[width=0.6\linewidth]{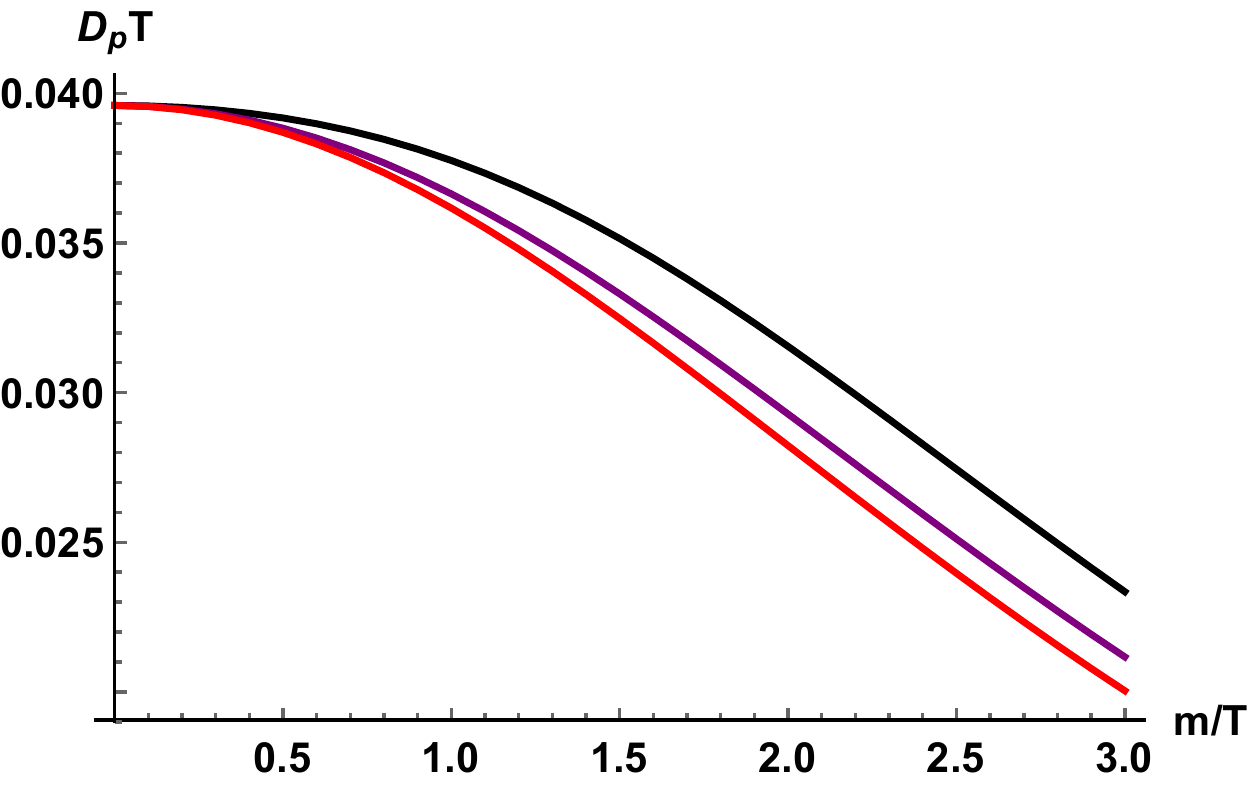}
    \caption{The diffusion constant of the sound mode, $D_p$, obtained from fitting the numerical data as a function of $m/T$ for various potentials $N=3,4,5$ (from black, top; to red, bottom).}
    \label{checkN}
\end{figure}

The equation for the diffusion constant can be considered as a sum of leading and sub-leading terms. The leading contribution, at $m/T \ll 1$, is
\begin{equation}\label{lead1}
    D_p = \frac12 \frac{\eta}{\chi_{\pi\pi}} + \dots,
\end{equation}
where the dots signify the sub-leading terms. The above expression reduces to the known result for sound damping, $D_p = \eta/(2sT)$, in the absence of SSB \cite{Policastro:2002tn,Baier:2007ix}. In figure \ref{phondifffig} we compare the behaviour of the dimensionless damping constant $D_p T$ with the numerical results computed for the potential with $N=3$, as functions of the symmetry breaking scale $m/T$. As expected, the leading order result \eqref{lead1} represents a good approximation only for small values of $m/T$.

Note that at large $m/T \rightarrow \infty$ the damping coefficient goes to zero and the phonons become purely propagating modes. In other words, at zero temperature no dissipation is present. The results are analogous for various $N>3$, see figure \ref{checkN}. The tendency is a smooth decrease when going to small temperature, which is consistent with the hypothesis that no dissipation can be at work at $T=0$; thus, in this limit our system can be considered a ``perfect elastic solid.''

\begin{figure}
    \centering
    \includegraphics[width=0.6\linewidth]{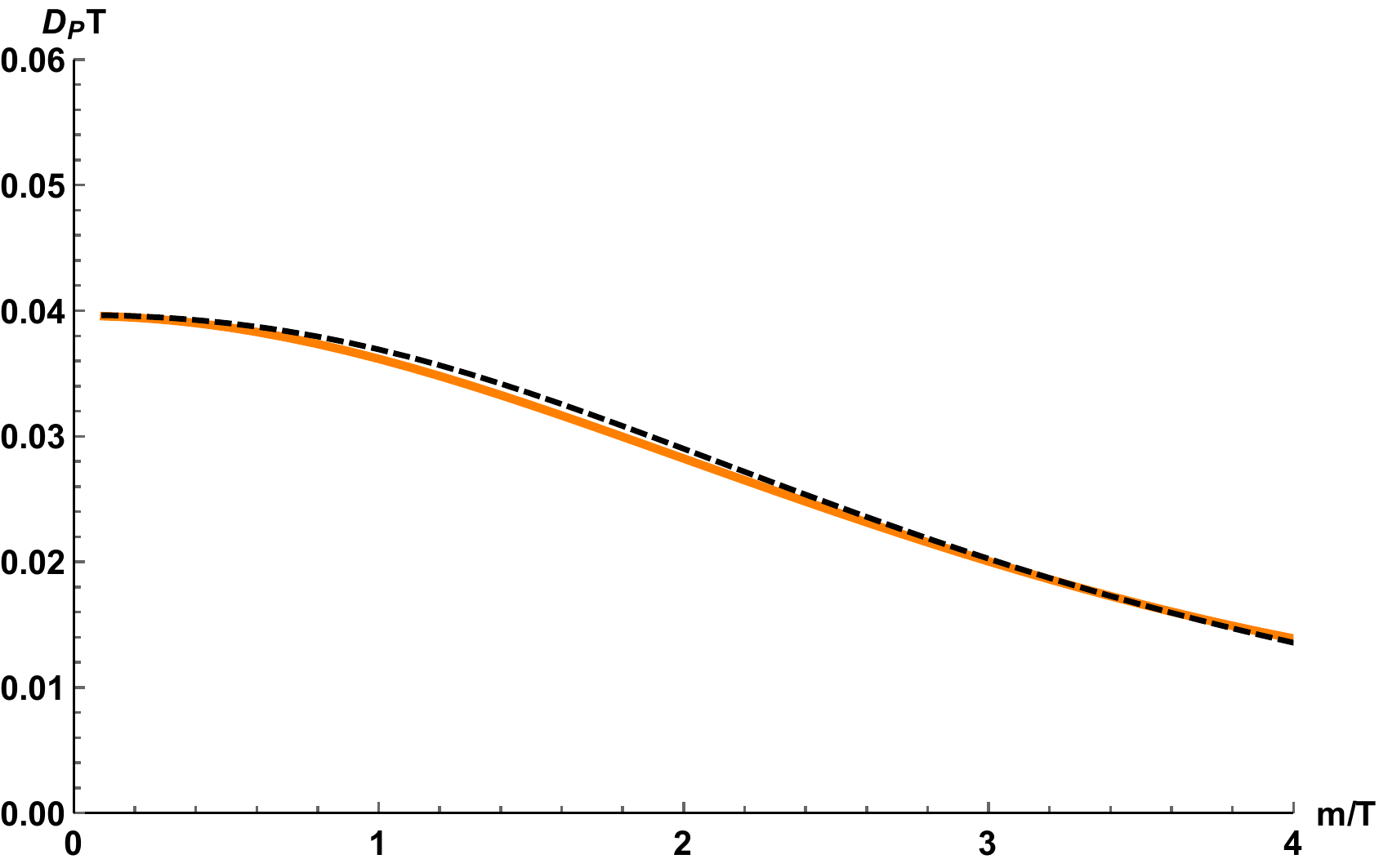}
    \caption{The comparison between the damping coefficient $D_p$ extracted from the numerical data (solid line) and the hydrodynamic formula \eqref{eq:diffusionModel}, for the specific $N=5$. We found similar results for several $N>5/2$. 
    }
    \label{figcheck1}
\end{figure}

We verified numerically, using the Kubo formulas \eqref{eq:kuboKappaMain} and \eqref{eq:kuboGammaMain}, that both coefficients $\bar{\kappa}_0$ and $\gamma_2$ are zero in our holographic model, which is in agreement with \cite{Kim:2016hzi}.\footnote{We thank Keun-Young Kim for explaining us this point.} Thus, the full expression for the diffusion constant of the sound modes reduces to 
\begin{equation}\label{eq:diffusionModel}
    D_p = \frac12 \frac{\eta}{\chi_{\pi\pi}} + \frac12 \,\frac{ (\kappa+G)^2\,\xi}{ \kappa + G + (\partial p/\partial \varepsilon)  \,\chi_{\pi\pi}}.
\end{equation}
In figure \ref{figcheck1} we plot this improved formula next to the numerical data, for $N=5$. It is evident that the complete formula, including the sub-leading terms in equation \eqref{eq:diffusionModel}, is now in very good agreement with the numerical data. 

\subsubsection{Diffusive Mode}

As mentioned above, spontaneous breaking of translational invariance gives rise to a non-propagating diffusive mode in the longitudinal sector.\footnote{Notice that such a mode does not appear when translations are broken explicitly as in \cite{Davison:2014lua}.} For an effective hydrodynamic description of this mode we refer the reader to \cite{PhysRevA.6.2401,PhysRevB.22.2514,Delacretaz:2017zxd} or to our appendix \ref{hydroapp}. This additional diffusive mode was first observed numerically in the context of holography \cite{Andrade:2017cnc} and later discussed in \cite{Baggioli:2019aqf}. In this subsection we will study this mode, in particular its diffusive constant $D_\Phi$. 

\begin{figure}
    \centering
    \includegraphics[width=0.6\linewidth]{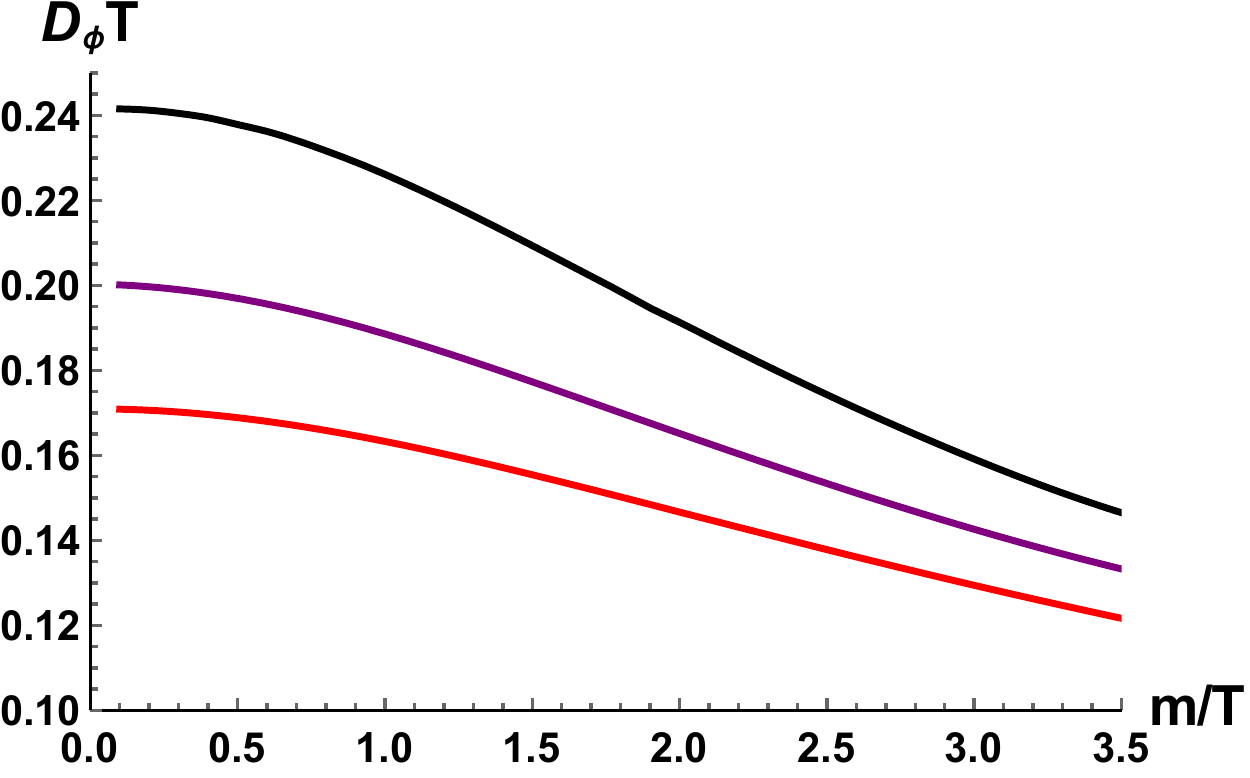}
    \caption{The numerical values for the Goldstone diffusion constant in function of $m/T$ for the potentials $N=3,4,5$ (from black, top; to red, bottom).}
    \label{figdiff}
\end{figure}

The numerical results for the dimensionless form of the diffusion constant are shown in figure \ref{figdiff}, for a range of values of the power $N$. We notice that the value of $D_\Phi$ is finite even at $m/T=0$, suggesting that its nature can be understood entirely in the limit when the Goldstone modes decouple.\footnote{Similar observations appeared in \cite{Amoretti:2018tzw}.} Moreover, $D_\Phi$ decreases monotonically when increasing the SSB parameter $m/T$.

\begin{figure}
    \centering
    \includegraphics[width=7.1cm]{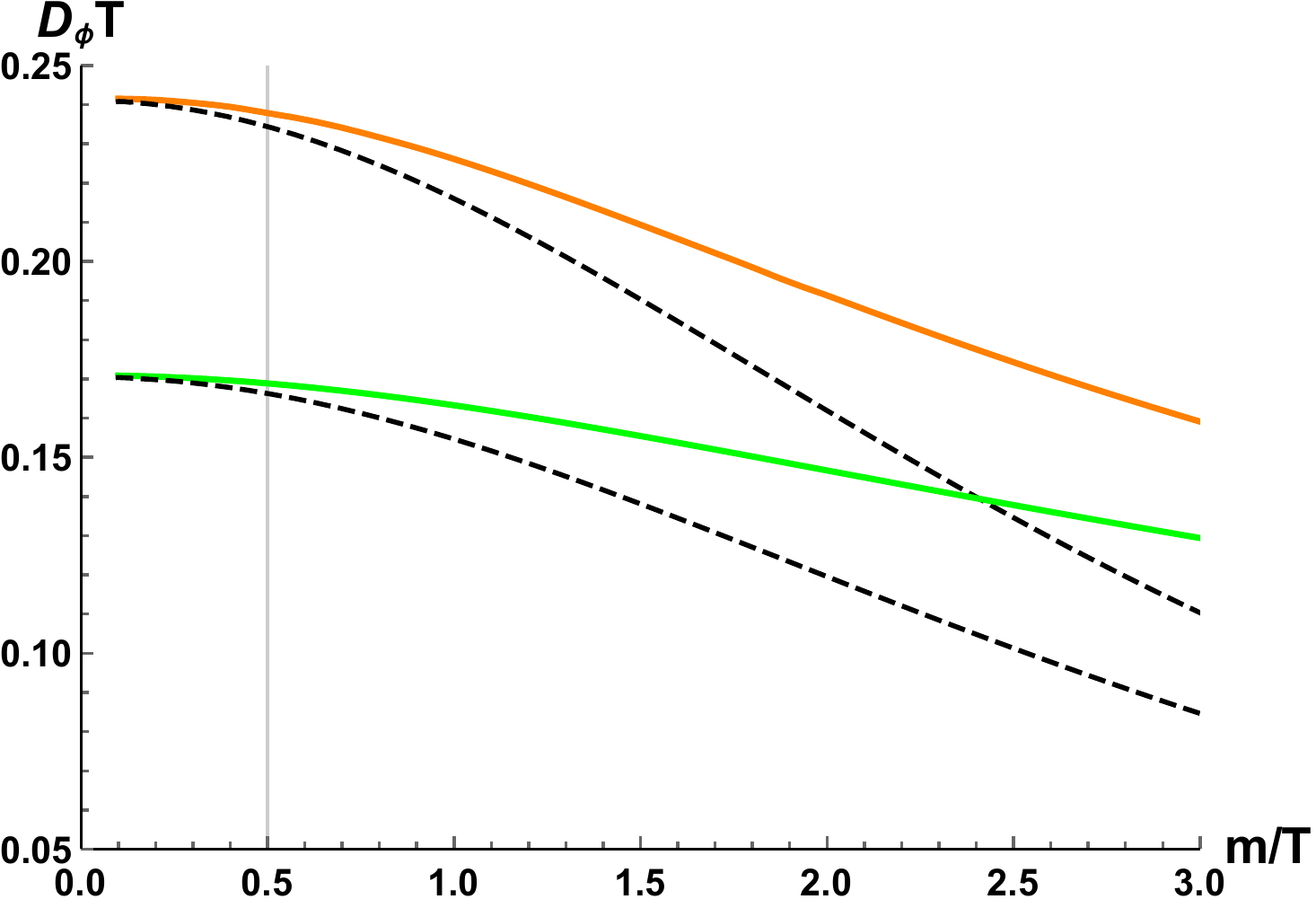}\quad\includegraphics[width=7.9cm]{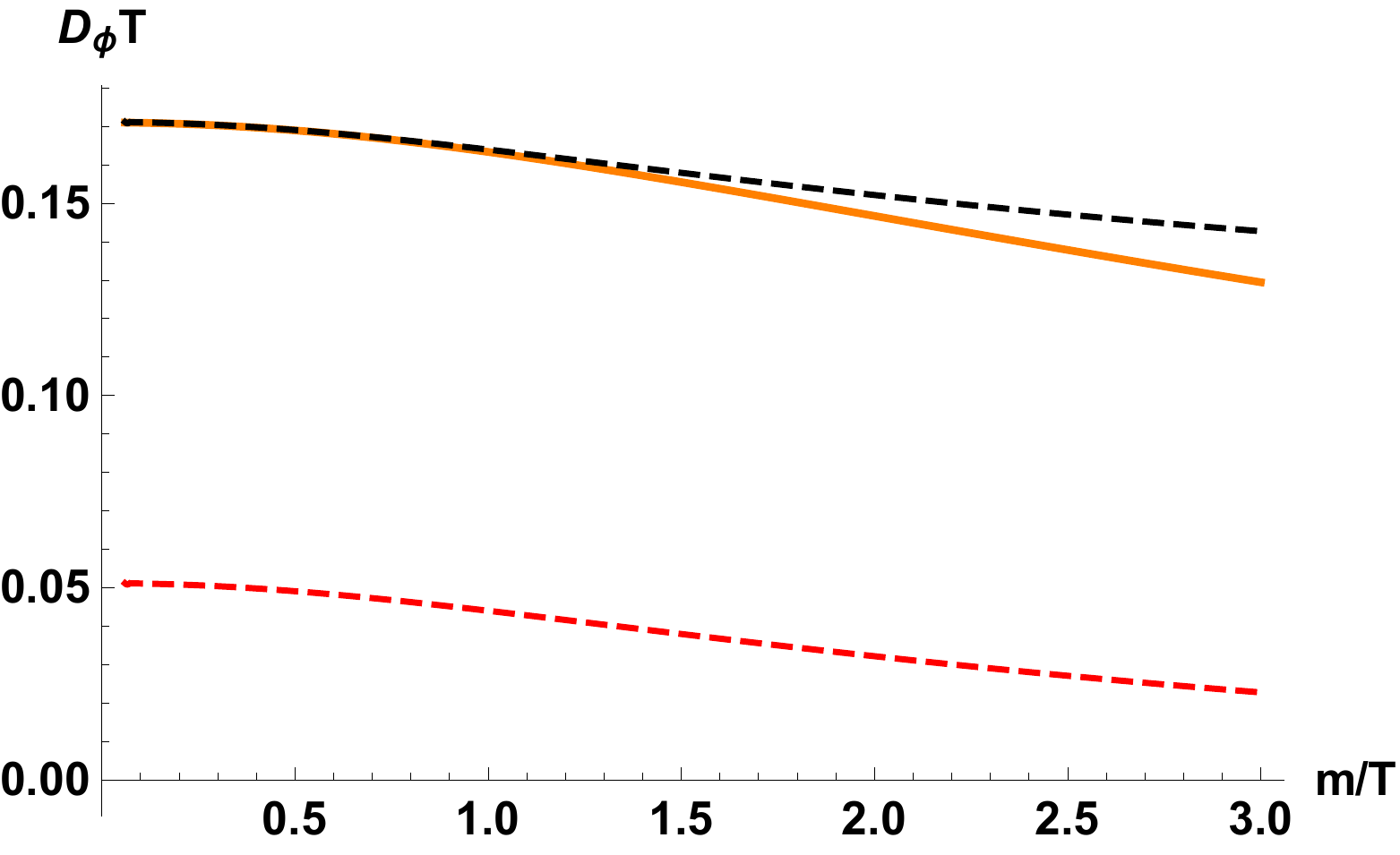}
    \caption{\textbf{Left: }The comparison between the numerical data for the diffusion constant $D_\Phi$ (solid lines) and the holographic formula \eqref{eq:diffusionAnalytic} (dashed lines). The different cases are for $N=3$ (upper, solid line) and $N=5$ (lower, solid line). \textbf{Right:} The comparison between the numerical value of the diffusion constant $D_\Phi$ for $N=5$ (solid line) and the predictions given by the hydrodynamic formula \eqref{eq:crystaldiffusionEnergy} (lower, dashed line). The upper dashed black line is the formula \eqref{eq:crystaldiffusionEnergy} with an additional constant shift. The agreement between the shifted hydrodynamic formula and the data is evident and valid until quite large values of $m/T$.}
    \label{fig:comp}
\end{figure}

In Appendix \ref{hydroapp} we derive the following analytic expression for the diffusion constant
\begin{equation}\label{eq:crystaldiffusionEnergy}
    D_\Phi = (\kappa + G) \,\frac{\bar{\kappa}_0 + \gamma_2\, \chi_{\pi\pi} + c_V \,(\partial p/\partial \varepsilon)  \,({T} \gamma_2 + \xi \,\chi_{\pi\pi}) }{c_V\, (\kappa + G + (\partial p/\partial \varepsilon)  \,\chi_{\pi\pi})}.
\end{equation}
In the case of $\bar{\kappa}_0=\gamma_2=0$, equation \eqref{eq:crystaldiffusionEnergy} displays a leading contribution
\begin{equation}\label{eq:diffusionLeading}
    D_\Phi= (\kappa+G) \,\xi + \dots\,,
\end{equation}
where the dots represent corrections. Surprisingly, we find that the complete hydrodynamic formula in expression \eqref{eq:crystaldiffusionEnergy} does not agree with the numerical quasi-normal mode analysis, which is evident in the plot in the right panel of figure \ref{fig:comp}.

Such a discrepancy between the numerical data and the analytic expression for $D_\Phi$ points towards a disagreement between the hydrodynamic predictions and the holographic results. Before continuing the analysis of this issue, let us first comment on the various parameters entering in the formula \eqref{eq:crystaldiffusionEnergy}:
\begin{enumerate}
    \item As stated in section \ref{sec:soundmode}, the coefficients $\bar{\kappa}_0$ and $\gamma_2$ are expected to vanish in absence of a finite charge density. Our numerical checks show that this is consistent in our model.
    \item The shear modulus $G$ has been used and tested in several previous works. More concretely, it has been tested using the dispersion relation of the transverse phonons in \cite{Alberte:2017oqx,Ammon:2019wci}.
    \item The bulk modulus $\kappa$ has been computed in the literature and is, additionally and successfully, tested in this manuscript using the speed of longitudinal sound.
    \item The dissipative coefficient $\xi$ has been used in \cite{Ammon:2019wci} to test the dispersion relation of the transverse modes, and its numerical value agrees perfectly with the horizon formula given in \cite{Amoretti:2018tzw}.
\end{enumerate}
In summary, we do not believe that the discrepancy between the hydrodynamic result and the holographic QNM data can be caused by a mistake in the computation of the parameters appearing in formula \eqref{eq:crystaldiffusionEnergy}.

We are able to gain some insight in to the puzzle by doing an analysis in the bulk. As stated in several works \cite{Baggioli:2019abx,Baggioli:2019aqf,Amoretti:2019cef,Amoretti:2018tzw,Donos:2019hpp,Donos:2019txg}, the diffusive Goldstone mode originates in the scalar sector and couples to the gravitational sector via the graviton mass $m$. In the limit of very small graviton mass, i.e. at values of $m/T\ll1$, one can approximate that the scalar and gravitational sectors decouple, hence the scalar fields can be treated as probes on a fixed gravitational background. In this decoupling limit the equation of motion for the Goldstone mode in the longitudnal sector reads
\begin{equation}
    \delta \phi_{\parallel}' \left(\frac{f'}{f}+\frac{2 i \omega }{f}+\frac{2 N}{u}-\frac{4}{u}\right)+\delta \phi_{\parallel} \left(-\frac{k^2 N}{f}+\frac{2 i N \omega }{u
   f}-\frac{4 i \omega }{u f}\right)+\delta \phi_{\parallel}''\,=\,0,\label{uno1}
\end{equation}
with the Schwarzschild emblackening factor given by
\begin{equation}
    f(u)\,=\,1\,-\,\left(\frac{u}{u_h}\right)^3.
\end{equation}
\begin{figure}
    \centering
    \includegraphics[width=0.6\linewidth]{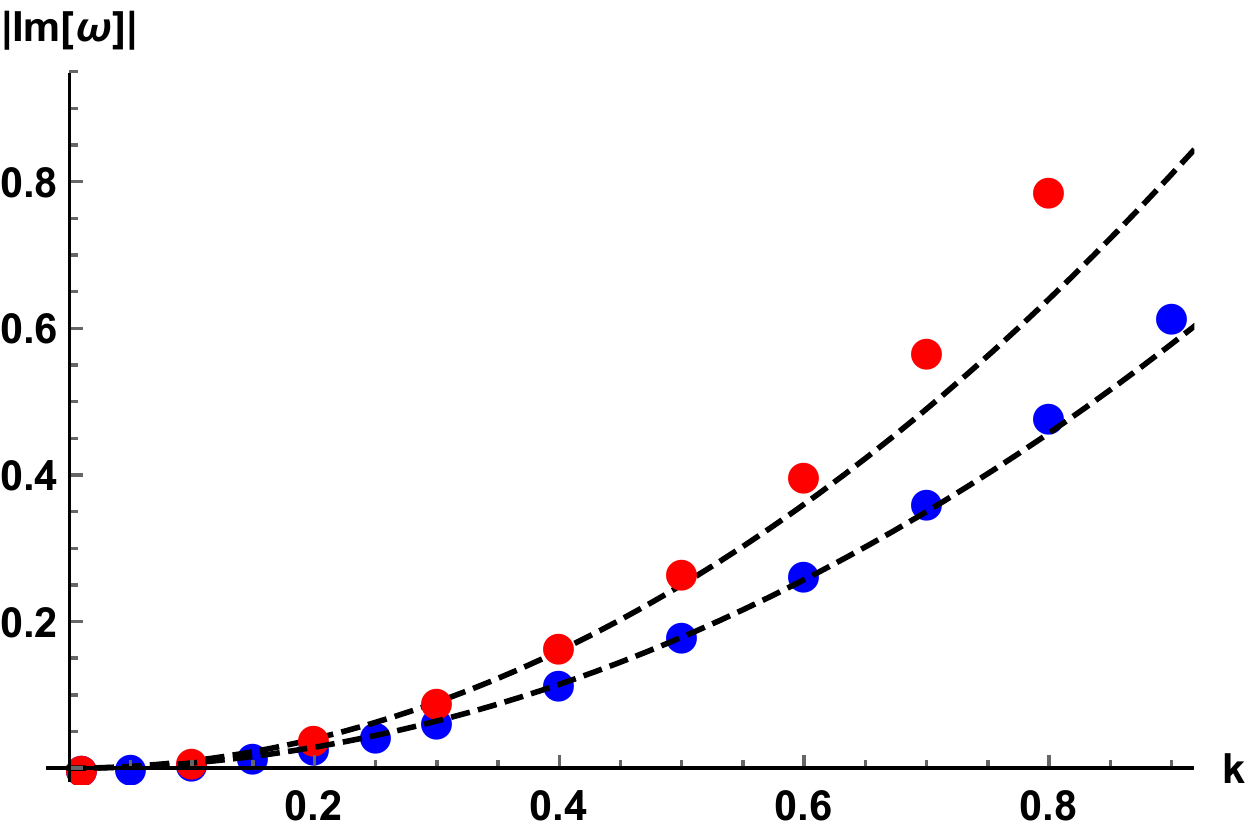}
    \caption{The diffusive mode in the decoupled scalar sector at $m/T=0$ for $N=3,5$ (red, upper; blue, lower). The dashed line is the analytic prediction.}
    \label{fignew}
\end{figure}
As expected, the quasi-normal modes calculated in this limit show the presence of a diffusive mode with dispersion relation
\begin{equation}
    \omega = -i \tilde{D}_\phi k^2 + \dots\, ,
\end{equation}
which is shown in figure \ref{fignew}. Given the simplicity of the decoupled regime, we are able to obtain the Green's function of the scalar operator analytically, using perturbative techniques. Implementing standard methods, we find the following expression for the diffusion constant
\begin{equation}\label{eq:diffusionAnalytic}
    \tilde{D}_\phi = \frac{N}{2N-3}\,u_h\,,
\end{equation}
where $N$ is the power in the potential $V(X)$ which defines the model. 

Continuing, in the limit of small graviton mass the constituents of the leading contribution to the hydrodynamic formula for $D_\Phi$, given by equation \eqref{eq:diffusionLeading}, can be written
\begin{equation}
    \xi\,=\,\frac{1}{N\,m^2}\,u_h^{4\,-\,2\,N},\quad  G\,=\,m^2\,\frac{N}{2N-3}\,u_h^{2\,N-\,3}\,+\,\mathcal{O}(m^4)\,,\quad \kappa\,=\,m^2\,\frac{N}{4N\,-\,6}\,u_h^{2N-3}.
\end{equation}
Using the above formulae we find the leading order of $D_\Phi$, at $m/T\ll1$, to be given by the formula
\begin{equation}\label{eq:diffusionLeadingAnalytic}
    D_\Phi\approx(\kappa\,+\,G)\,\xi\,=\,\frac{3}{2}\,\frac{1}{2N-3}\,u_h\,+\,\mathcal{O}(m^2).
\end{equation}
Comparing \eqref{eq:diffusionAnalytic} and \eqref{eq:diffusionLeadingAnalytic}, it is evident that the leading contribution to the hydrodynamic formula for the diffusion constant of the diffusive mode does not match the corresponding quantity extracted from holography, even at $m/T\ll1$. Moreover we can see that
\begin{equation}\label{eq:diffusionRelation}
    \tilde{D}_\phi = \frac{2}{3} N (\kappa + G)\xi = N G \xi = N D_\perp,
\end{equation}
where we in the second equality use the identity $(\kappa + G)/G=3/2$; and $D_\perp=G \xi$ is the diffusion constant of the transverse diffusive mode, which has been successfully tested in \cite{Amoretti:2018tzw,Ammon:2019wci}. The above relation can also be seen directly by comparing the equation of motion \eqref{uno1} to its transverse counterpart
\begin{equation}
    \delta \phi_{\perp}' \left(\frac{f'}{f}+\frac{2 i \omega }{f}+\frac{2 N}{u}-\frac{4}{u}\right)+\delta \phi_{\perp} \left(-\frac{k^2}{f}+\frac{2 i N \omega }{u
   f}-\frac{4 i \omega }{u f}\right)+\delta \phi_{\perp}''\,=\,0.\label{uno2}
\end{equation}
The plot shown in the left panel of figure \ref{fig:comp} indicates that the formula \eqref{eq:diffusionRelation} is in good agreement with the numerical data in the limit of soft SSB, roughly $m/T \lesssim 0.5$. 

 It is important to note that a missing factor of $N$ is not able to fix the disagreement between hydrodynamics and the holographic results at arbitrary values of $m/T$. This is clear from the left panel of figure \ref{fig:comp}, where the discrepancy is still present at values of $m/T \gtrsim 1$. Taking the difference between \eqref{eq:diffusionLeadingAnalytic} and \eqref{eq:diffusionLeading} we find
 \begin{equation}
     \tilde{D}_\phi = (\kappa + G) \xi + \frac{1}{2}\,u_h,
 \end{equation}
 which means we at leading order in contribution have the relation
  \begin{equation}
     \tilde{D}_\phi = D_\Phi + \frac{1}{2}\,u_h.
 \end{equation}
 In terms of dimensionless quantities the above relation becomes
 \begin{equation}
     \tilde{D}_\phi\,T = D_\Phi\,T\, +\, \frac{3}{8\,\pi}.
 \end{equation}
Surprisingly, we have found that this discrepancy is independent of the choice of potential, at least for small $m/T$. The predictions of hydrodynamics combined with such a shift reproduces the numerical values of the diffusion constant very well  for potentials of the form $V(X)=X^N$, even until large values of $m/T$, see the right panel of figure  \ref{fig:comp}.

We are currently unable to find a precise hydrodynamic resolution for the mismatch described above. Assuming the correctness of our computations, we cannot discard the possibility that the hydrodynamic description is missing some effect encoded in a novel transport coefficient. At the same time, it is not guaranteed that the holographic models considered in this work are the exact duals of a system with spontaneously broken translations, as those described by hydrodynamics. One possible reason for the discrepancy is that the susceptibility matrix defined in eq.\eqref{eq:suscebtibilitymatrixFULL} is not as simple and in particular it might contain off diagonal terms. A preliminary analysis shows that such off-diagonal terms will not appear in our model and that they will not affect the results at $m=0$. A more detailed investigation in this direction is needed.

A last possibility is related to the fact that our vacuum does not minimize the free energy. In other words, our solution is not thermodynamically favourable and hence it can been seen as an excited state of the dual field theory. The situation is nevertheless more subtle, because our theory can not be constructed around the would-be thermodynamic favourable vacuum $\phi^I=0$. More precisely, the theory we consider is strongly coupled around such a background and the results are therefore not trustable. In principle this should not affect the quantities we are discussing and indeed for all of them beside the Goldstone diffusion constant we find very good agreement. In summary, our results are surprising given the large amount of verifications in this direction.

\section{Quasi-Normal Modes Beyond Hydrodynamics}\label{sec:beyond}
In the previous section we investigated hydrodynamic modes, i.e. poles of the retarded Green's function located at $\omega/T \ll 1, k/T \ll 1$. Hence, these modes determine the late time and long distance behaviour of the system. 

Using holographic techniques we can study further, so-called non-hydrodynamic, poles of the Green's functions. These poles correspond to higher quasi-normal modes which we determine numerically. In particular we are interested in the behaviour of the higher quasi-normal modes as functions of the SSB scale $m/T$, and momentum $k/T$.
\begin{figure}
\centering
\includegraphics[width=6cm]{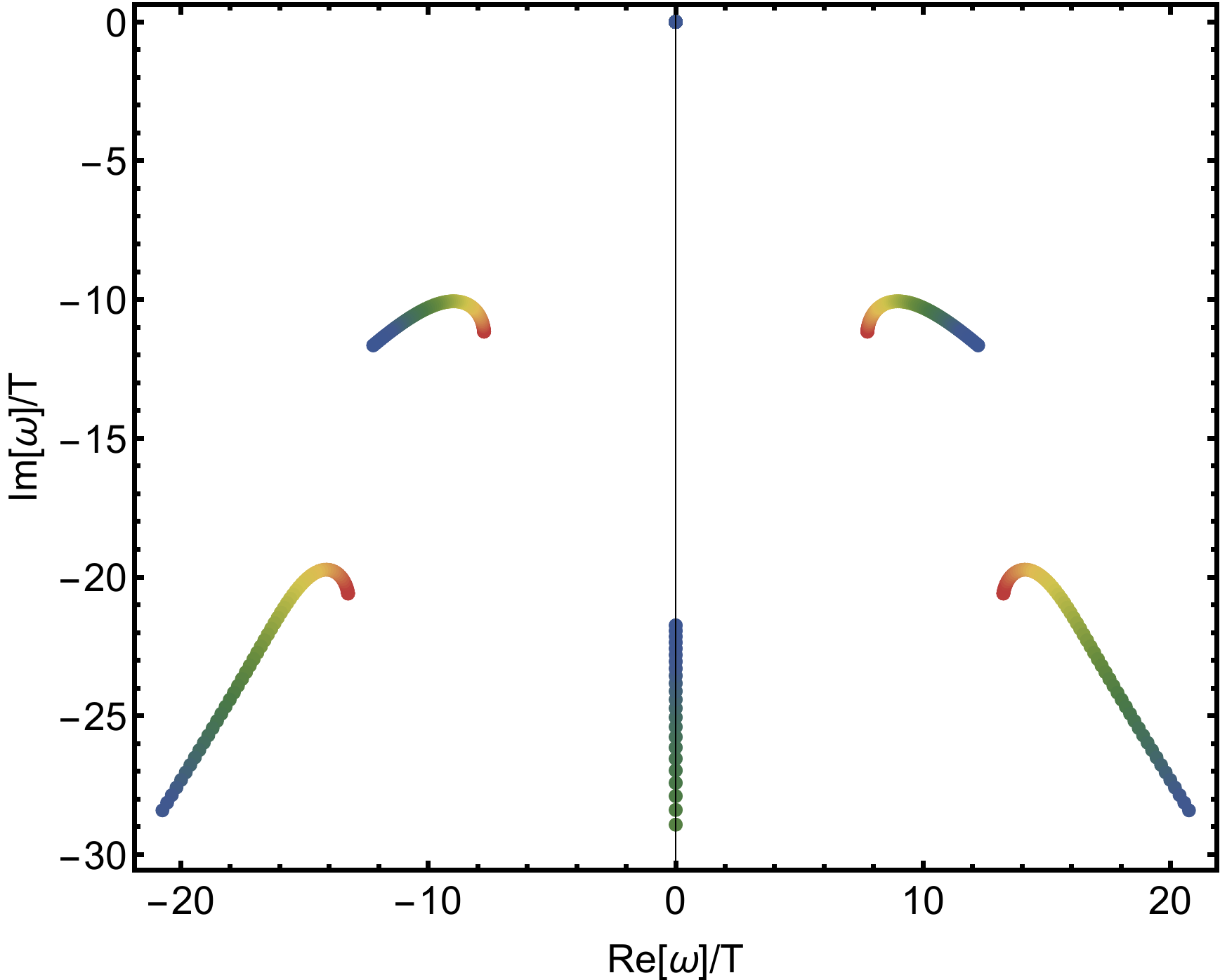}
\quad
\includegraphics[width=6cm]{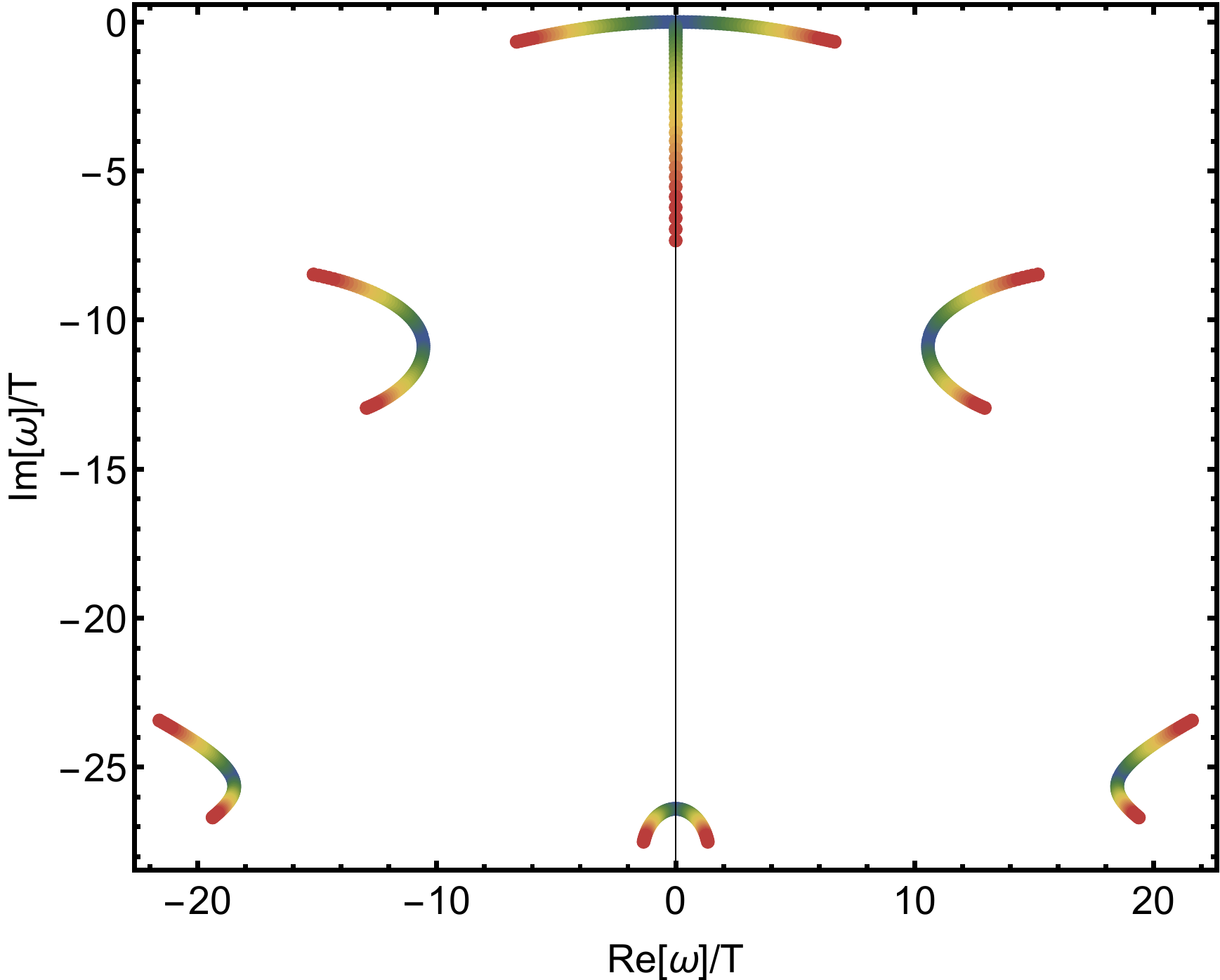}
\caption{Spectrum of higher QNMs for the potential $V(X)=X^3$. \textbf{Left:} The QNM spectrum for vanishing momentum as a function of the dimensionless (inverse) temperature $m/T \in [0, 6.5].$ Blue dots denote low temperatures while red dots refer to high temperatures. \textbf{Right:} The QNM spectrum for fixed temperature $m/T=0.179$ as a function of the dimensionless momentum $k/T \in [0.186,7.45].$ \label{pic:moving}}
\end{figure}

The results for the potential $V(X)=X^3$ are shown in figure \ref{pic:moving}. The left panel of the figure displays the tower of quasi-normal modes at zero momentum while dialing the dimensionless parameter $m/T$. For any given ratio $m/T$ there is a clear gap in the system. Moreover, as the mass $m/T$ is increased all the higher non-hydrodynamic modes move away from the origin of the complex plane and the imaginary part of the frequency grows. Hence the late time and long distance behaviour is determined by the three hydrodynamic modes which we investigated in the previous section. The other poles do not give rise to quasi-particles due to their short lifetime.

\begin{figure}
    \centering
    \includegraphics[width=6cm]{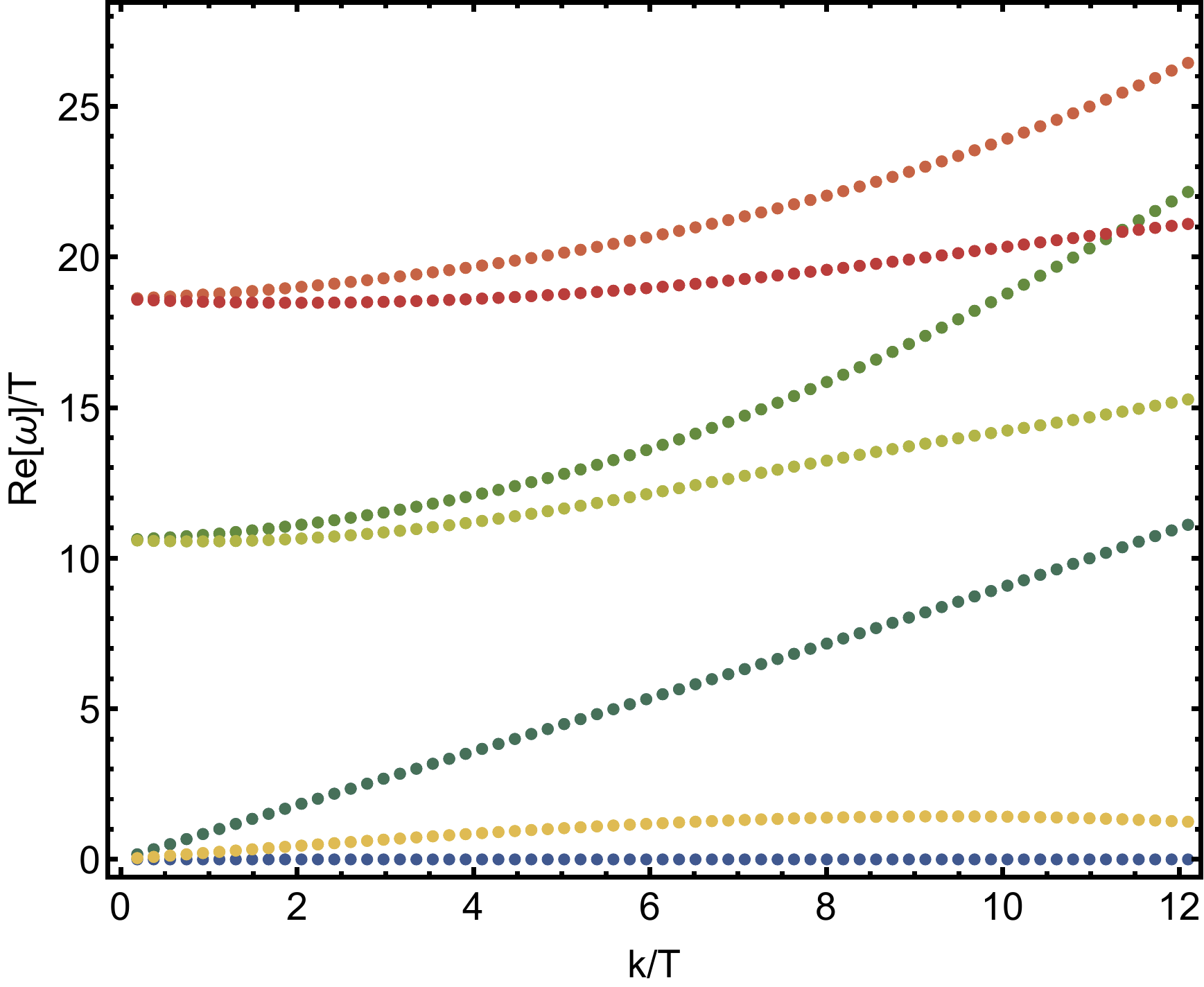}
    \quad
    \includegraphics[width=6cm]{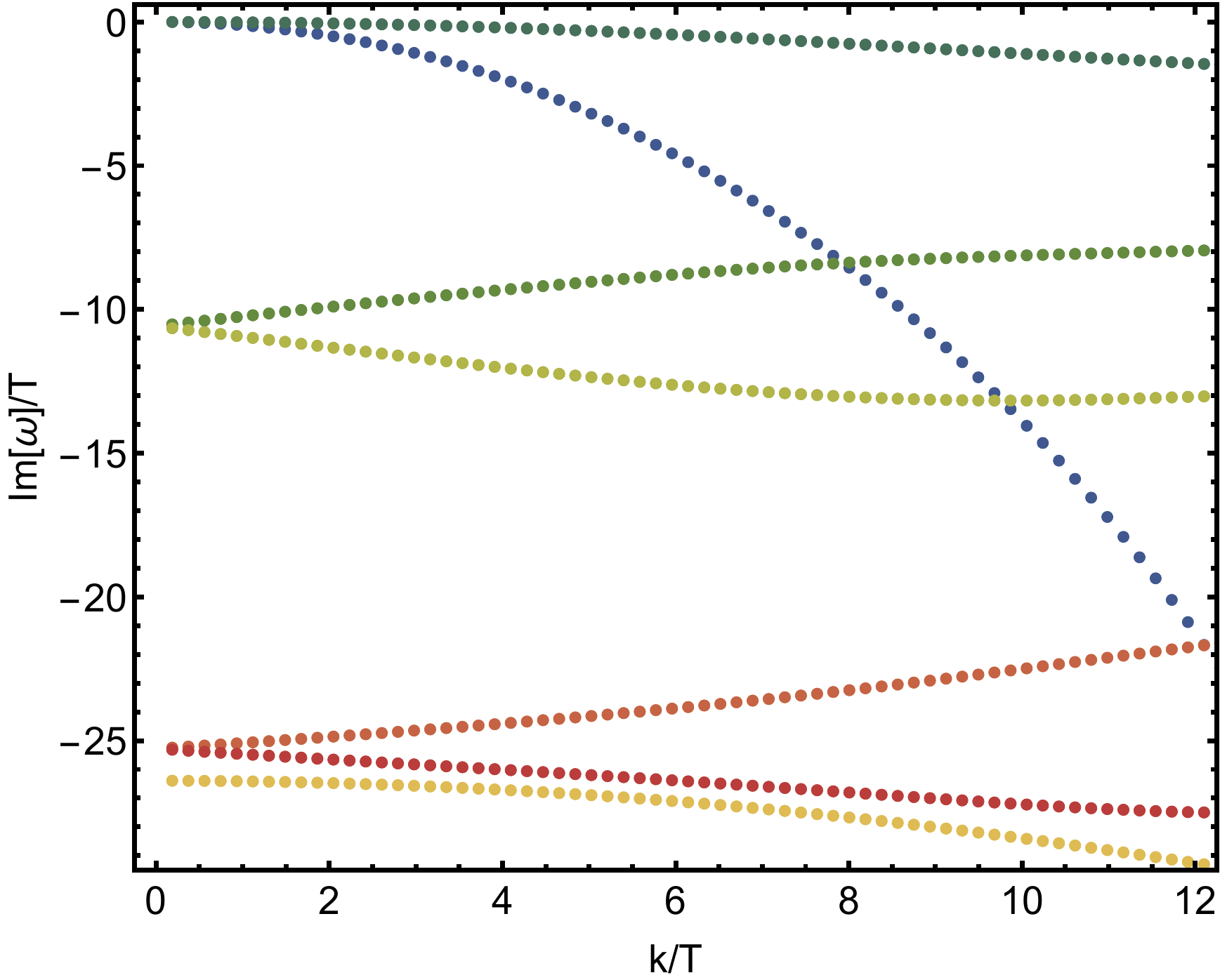}
    
    \vspace{0.3cm}
    
    \includegraphics[width=6cm]{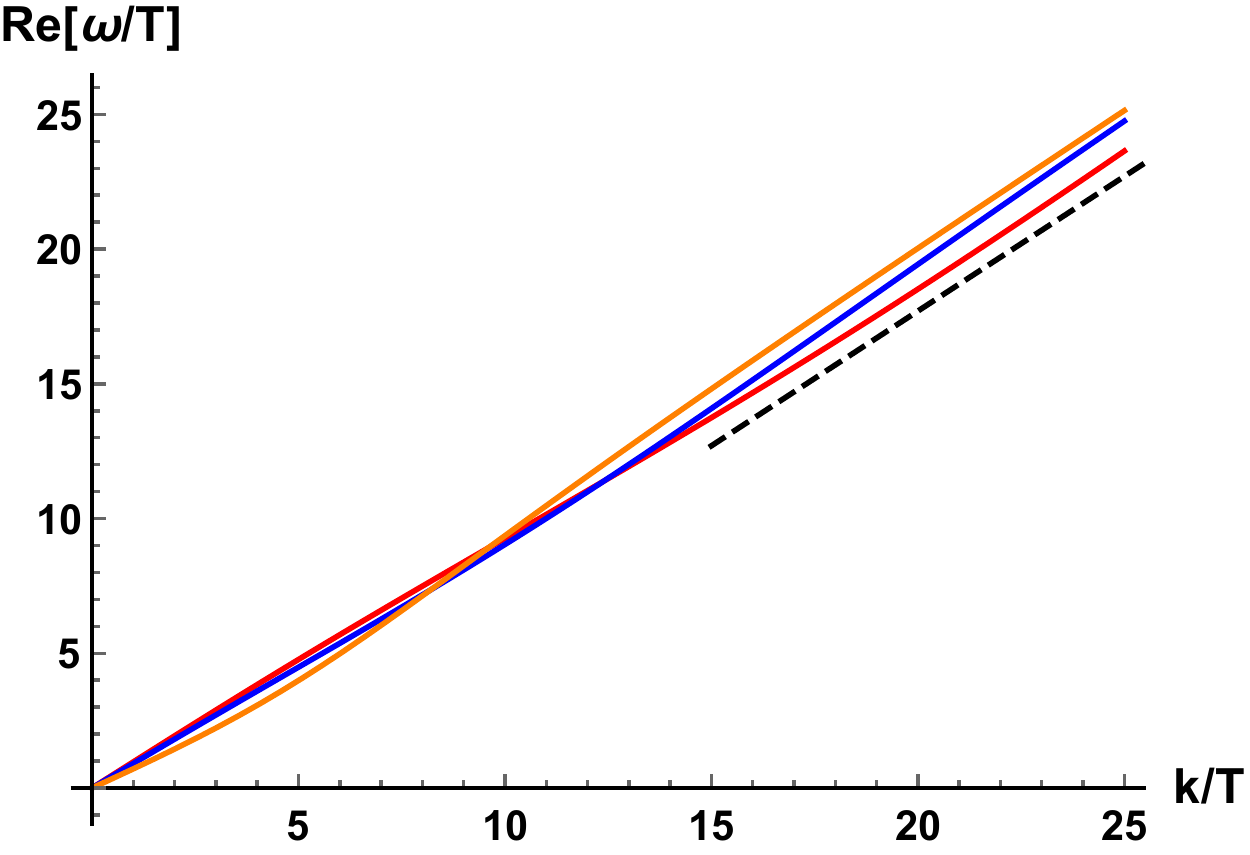}
    \quad
    \includegraphics[width=6cm]{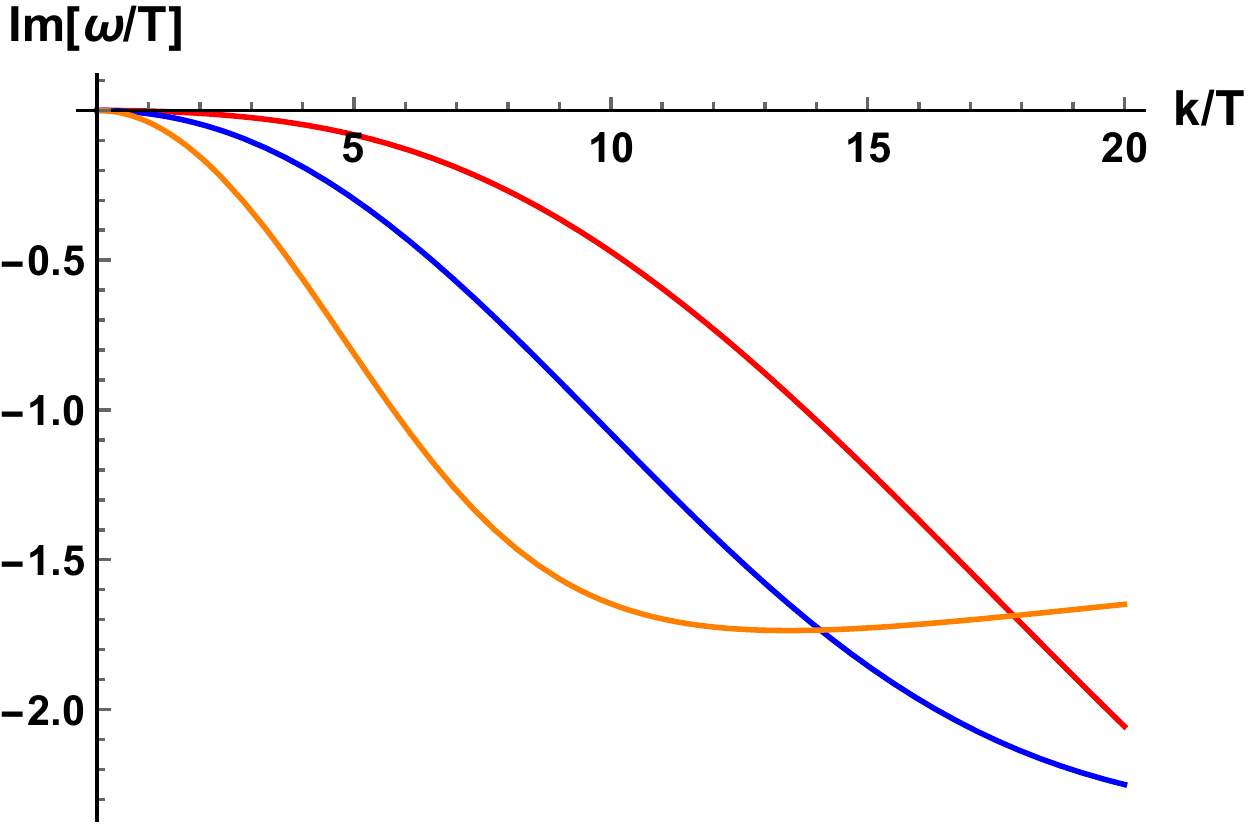}
    \caption{\textbf{Top Panel:} The full spectrum of QNMs for fixed temperature $m/T=0.179$. \textbf{Bottom Panel:} The dispersion relation of the longitudinal sound mode beyond the hydrodynamic limit. In the hydrodynamic limit, $k/T\ll 1$, the behaviour is described by (\ref{mode1}). At large momenta, i.e. for $k/T\gg 1$, the real part of the frequency asymptotes to the UV dispersion relation $\textrm{Re}(\omega)=\pm k$ indicated by the dashed black line. The imaginary part of the frequency is small compared to the real part even for large momenta.}
    \label{fig:beyond}
\end{figure}

In the right panel of figure \ref{pic:moving}  we show the quasi-normal mode spectrum as a function of the momentum $k$. In addition, the more refined results for the first seven modes are shown in figure \ref{fig:beyond}. Note that the dynamics of the higher quasi-normal modes as a function of the momentum appears to be quite complicated. Especially noticeable is the large imaginary part of the QNM depicted in blue in figure \ref{fig:beyond} (corresponding to the diffusive mode in the hydrodynamic regime) upon increasing $k/T$.

Finally, we analyse the large momentum behaviour of the quasi-normal modes corresponding to the two hydrodynamic sound modes. In particular, we consider the case $k/T \gg 1$ which clearly exceeds the hydrodynamic regime. As discussed above, at small momenta $k/T \ll 1$ the expected dispersion relation is given by \eqref{mode1}. On the contrary at very large momenta we obtain the dispersion relation of a sound mode within the relativistic UV AdS fixed point. In this limit, the real part of the dispersion relation is given by $\textrm{Re}(\omega)=\pm \,k$. 

In contrast to the sound mode in the transverse sector (see \cite{Ammon:2019wci}), the real part of the frequency of the sound mode never becomes zero; in fact, as shown in \ref{fig:beyond} the real part of the dispersion relation of the sound mode interpolates smoothly between $\textrm{Re}(\omega) = \pm c_L\, k$ in the hydrodynamic limit and $\textrm{Re}(\omega)=\pm\, k$ in the large momentum limit. 

\section{Conclusions}
\label{sec: conclusions}
In this paper we studied the longitudinal quasi-normal mode spectrum of a simple holographic system with spontaneously broken translational symmetry, at zero charge density. We successfully identify the three hydrodynamic modes: the pair of longitudinal sound modes and a diffusion mode.  We compared the numerical results obtained from the holographic model with the predictions of hydrodynamics given in \cite{Delacretaz:2017zxd} and in our appendix \ref{hydroapp}. 

The numerical results for the propagation speed and the attenuation constant of the sound modes are in agreement with the hydrodynamic formulas at any strength of the SSB, at arbitrary graviton mass. Surprisingly, we find disagreement between the diffusion constant of the diffusion mode obtained numerically and the formula given by hydrodynamics. The discrepancy is corroborated with analytic methods in the decoupling limit and it is present even at zero graviton mass. At this stage we are not able to identify the reason behind this puzzle. 

Some comments are in order. Mistakes present in our computations are of course a possibility. A second, more interesting, option is that the hydrodynamic framework considered is missing some effect. A lacking transport coefficient, for example, could potentially explain the discrepancy we observe. A last scenario is that the holographic model discussed in this work is not the gravity dual of the type of systems described by the hydrodynamic picture, i.e. a dissipative and conformal system with spontaneously broken translational invariance. Needless to say, finding the reason for this discrepancy is certainly the most important open question that our manuscript poses. It would be interesting to perform the same check in similar holographic models \cite{Andrade:2017cnc,Grozdanov:2018ewh,Amoretti:2018tzw,Baggioli:2018bfa}.

As a final remark, it would be beneficial to gain a better understanding regarding the propagation of sound and the elastic properties of quantum critical materials, both from EFT methods \cite{Alberte:2018doe} and holographic methods \cite{oriol}. There are several recent hints suggesting that phonons, viscoelasticity and glassy behaviours may play an important role in quantum critical systems \cite{ishii2019glass} and High-Tc superconductors \cite{setty2019glass}. Moreover, electron-phonon coupling might play an important role in the high-Tc puzzle \cite{He62}.

In conclusion, we have in this paper continued the study of simple holographic massive gravity models with spontaneously broken translational symmetry. We have tested several predictions obtained from the hydrodynamic theory of such systems. Finally, we provided food for thought in the form of a discrepancy between the hydrodynamic and holographic theories concerning the dispersion relation of the diffusive mode. In this case, we would be very happy to be proven wrong, and even more happy to find a novel and interesting physical reason behind the mismatch we observed.

\section*{Acknowledgements}
We thank Mike Blake, Alex Buchel, Markus Garbiso, Blaise Gouteraux, Saso Grozdanov,  Matthias Kaminski, Keun-Young Kim, Napat Poovuttikul and Vaios Ziogas for useful discussions and comments about this work and the topics considered. We are particularly grateful to: Daniel Arean for sharing with us precious information; Amadeo Jimenez Alba for collaboration on closely related topics and for never-ending critical discussions; Oriol Pujolas and Victor Cancer Castillo for collaboration on closely related topics and for sharing with us unpublished results. Finally, we would like to sincerely thank Blaise Gouteraux for several discussions and suggestions.

MA is funded by the Deutsche Forschungsgemeinschaft (DFG, German Research  Foundation) -- 406235073. 
MB acknowledges the support of the Spanish MINECO’s “Centro de Excelencia Severo Ochoa” Programme under grant SEV-2012-0249. SG gratefully acknowledges financial support by the  \textit{Fulbright Visiting Scholar Program}, which is sponsored by the US Department of State and the German-American Fulbright Commission in 2018 and  by the DAAD (German Academic Exchange Service) for a \textit{Jahresstipendium f\"ur Doktorandinnen und Doktoranden} in 2019.

MA would like to thank the Erwin Schr\"odinger International Institute for Mathematics and Physics as well as NORDITA for the hospitality during the completion of this work. MB would like to thank MIT, Perimeter Institute, University of Montreal, University of Alabama and Universidad Autonoma de Mexico for the warm hospitality during the completion of this work.
MB would like to thank Marianna Siouti for the unconditional support.

\appendix
\section{Equations of Motion}\label{app1}
In this work we determine the spectrum of quasi-normal modes in the longitudinal sector. Choosing the momentum along the $y$-direction,  the relevant time-dependent fluctuations are $\delta\phi_2, h_{tt}, h_{ty} ,h_{yu}, h_{xx}, h_{uu}, h_{yy}$ and $h_{tu}$. 
We choose to work in radial gauge i.e. $h_{\mu u}=0$, with $\mu\in\{t,u,x,y,z\}$. Since we work in in-falling Eddington-Finkelstein coordinates, the fluctuations already satisfy in-going horizon conditions. The Fourier transformed equations of motions to first order in the fluctuations read 
\begin{align}
0&={i} k\, u\, h_{x,a}-\text{i} k\, (N-1)\, u\, h_{x,s}+u\, h_{ty}'+2\, (N-2) h_{ty}+ \left(-k^2 N u+2 \text{i} N \omega -4 \text{i}\, \omega \right)\delta \phi_2\nonumber\\
&\quad+ \left(u f'+2 \text{N} f-4 f+2 \text{i} \,u\,\omega \right)\delta \phi_2'+u f \delta \phi_2''\\[0.2cm]
0&=u^2 f f' h_{x,s}'+u\, \omega  \left(i u f'-2 i f+2 u\, \omega \right) h_{x,s}+2 m^2 (N-1) N f u^{2 N} \left(h_{x,s}-i k \,\delta \phi_2\right)\nonumber\\
&\quad-2 u f^2\, h_{x,s}'+k\, u\, h_{ty} \left(i u f'-2 i f+2 u\, \omega \right)+4 u f\, h_{tt}'-u^2 f h_{tt}''\nonumber\\
&\quad+\left(-u^2 f''+4 u f'-12 f+k^2 u^2+2 m^2 (N-1) u^{2 N}-2 i u \,\omega +6\right)h_{tt}\\[0.2cm]
0&=-k\, u^2 \omega\,  h_{x,a}-k u^2\, \omega\,  h_{x,s}+2 h_{ty} \left(u f'-3 f+m^2 (N-1) u^{2 N}+3\right)\nonumber\\
&\quad-i k\, u^2 h_{tt}'+2 i k\, u\, h_{tt}+2 i\, m^2 N \omega\,  u^{2 N} \delta \phi_2+\left(2 u f-i u^2 \omega \right) h_{ty}'-u^2 f h_{ty}''\\[0.2cm]
0&=h_{x,s} \left(2 u f'\!-\!6 f+k^2\, u^2+2 m^2 (N\!-\!1) u^{2 N}+4 i u\, \omega\! +\!6\right)+u \left(-u f'+4 f-2 i u\, \omega \right) h_{x,s}'\nonumber\\
&\quad\!+\!k^2 u^2 h_{x,a}+6 h_{tt}\!-i k\, u^2\, h_{ty}'-2 u h_{tt}'+4 i k\, u\, h_{ty}-2 i k\,m^2 \,N u^{2N}\delta \phi_2\! -u^2\, f\, h_{x,s}''\\[0.2cm]
0&=2 \left(u f'\!-\!3 f+m^2 (N\!-\!1) u^{2N}+i u\, \omega\! +\!3\right)h_{x,a}\! +u \left(-u f'\!+\!2 f\!-\!2 i u\, \omega \right) h_{x,a}'-u^2\, f h_{x,a}''\nonumber\\
&\quad+i k\, u^2\, h_{ty}'-2 i k\, u\, h_{ty}+2 i k \,m^2\, N\, u^{2 N}\, \delta \phi_2\\
0&=u \left(\left(-2 f+u \left(f'+2 i\, \omega \right)\right) h_{x,s}'-2 i\, \omega\,  h_{x,s}-u\, h_{tt}''+4 h_{tt}'+i\, k\, u\, h_{ty}'-2 i k\, h_{ty}\right)\nonumber\\
&\quad+2 m^2 \,(N\!-\!1) N\, u^{2N} \left(h_{x,s}-i k\, \delta \phi_2\right)-6\, h_{tt}\\[0.2cm]
0&=u \left(k\, u \left(h_{x,a}'+h_{x,s}'\right)-i\, u\, h_{ty}''+2 i\, h_{ty}'\right)-2 i\, m^2 N\, u^{2 N} \,\delta \phi_2'\\
0&=h_{x,s}''~,\label{eq:lcon}
\end{align}
where we introduced the short-hand notations $h_{x,s}$ and $h_{x,a}$ for the symmetric combination $h_{x,s}=1/2\,(h_{xx}+h_{yy})$ and the anti-symmetric combination $h_{x,a}=1/2\,(h_{xx}-h_{yy})$, respectively. In terms of these combinations it is straightforward to solve equation \eqref{eq:lcon} by integrating twice, $h_{x,s}=c_1+c_2\,h_{x,s} u$. In order to determine the integration constants, we compare the solution to the near the boundary expansion at the conformal boundary given by
\begin{equation}
 h_{x,s}(u\to0)=h^{\text{source}}_{x,s}+\ldots+u^3\,h^{\text{vev}}_{x,s}  +\ldots\,.
\end{equation}
In the context of the QNM computations we do not allow for sources of fluctuations $c_1$ and $c_2$ have to vanish leading to the trivial solution $h_{x,s}=0$.\footnote{In order to compute the Kubo formulas we have to source some of the metric functions; in cases where we source $T_{xx}$ or $T_{yy}$  we can not set $ h_{x,s}$ to zero.} In this case, the equations of motion simplify to
\begin{align}
0=&\,\text{i} k\, u\, h_{x,a}+u\, h_{ty}'+2\, (N-2) h_{ty}+ \left(-k^2 N u+2 \text{i} N \omega -4 \text{i}\, \omega \right)\delta \phi_2\nonumber\\&+ \left(u f'+2 \text{N} f-4 f+2 \text{i} \,u\,\omega \right)\delta \phi_2'+u f \delta \phi_2''\label{eq:EOM1}\\[0.2cm]
0=&\,2 m^2 (N\!-\!1) N f u^{2 N} \left(-i k \,\delta \phi_2\right)+k\, u\, h_{ty} \left(i u f'-2 i f+2 u\, \omega \right)+4 u f\, h_{tt}'-u^2 f h_{tt}''\nonumber\\&+\left(-u^2 f''+4 u f'-12 f+k^2 u^2+2 m^2 (N-1) u^{2 N}-2 i u \,\omega +6\right)h_{tt}\\[0.2cm]
0=&-k\, u^2 \omega\,  h_{x,a}+2 h_{ty} \left(u f'-3 f+m^2 (N-1) u^{2 N}+3\right)\nonumber\\&-i k\, u^2 h_{tt}'+2 i k\, u\, h_{tt}+2 i\, m^2 N \omega\,  u^{2 N} \delta \phi_2+\left(2 u f-i u^2 \omega \right) h_{ty}'-u^2 f h_{ty}''\\[0.2cm]
0=&\,k^2 u^2 h_{x,a}+6 h_{tt}\!-i k\, u^2\, h_{ty}'-2 u h_{tt}'+4 i k\, u\, h_{ty}-2 i k\,m^2 \,N u^{2N}\delta \phi_2\\[0.2cm]
0=&\,2 \left(u f'\!-\!3 f+m^2 (N\!-\!1) u^{2N}+i u\, \omega\! +\!3\right)h_{x,a}\! +u \left(-u f'\!+\!2 f\!-\!2 i u\, \omega \right) h_{x,a}'\nonumber\\&+i k\, u^2\, h_{ty}'-2 i k\, u\, h_{ty}+2 i k \,m^2\, N\, u^{2 N}\, \delta \phi_2-u^2\, f h_{x,a}''\\[0.2cm]
0=&\,u \left(-u\, h_{tt}''+4 h_{tt}'+i\, k\, u\, h_{ty}'-2 i k\, h_{ty}\right)-2i\, k\, m^2 \,(N\!-\!1) N\, u^{2N}\,\delta \phi_2-6\, h_{tt}\\[0.2cm]
0=&\,u \left(k\, u\, h_{x,a}'-i\, u\, h_{ty}''+2 i\, h_{ty}'\right)-2 i\, m^2 N\, u^{2 N} \,\delta \phi_2'~.\label{eq:EOMlast}
\end{align}

\section{Numerical Techniques}\label{app2}
\subsection*{Quasi-Normal Modes}
In this section, we briefly review the numerical methods for the quasi-normal modes used throughout this work. For numerical convenience, we collect \eqref{eq:EOM1}-\eqref{eq:EOMlast} in a system of equations of the form \begin{equation}
    (a-\omega \bm{B})\,\bm{x}=0, \label{equation:eigen}
\end{equation}
with $\bm{x}= \{\delta\phi_2,h_{tt},h_{ty},h_{x,a}\}$ and $a,\bm{B}$ being differential operators. The quasi-normal modes are the complex frequencies $\omega_n$ to which $\bm{x}_n$ is a regular solution to equation \eqref{equation:eigen} (and fulfills the constraint equations).
For given $N,k,$ and $m$, we may solve the eigenvalue problem by means a of pseudo-spectral method on a Gau\ss-Lobatto grid, as explained in \cite{Grieninger:2017jxz,Ammon:2016fru}. Note, that the sources of the fluctuations have to vanish which we implement by a suitable redefinition of the fields.

\begin{figure}
\centering
\includegraphics[width=6cm]{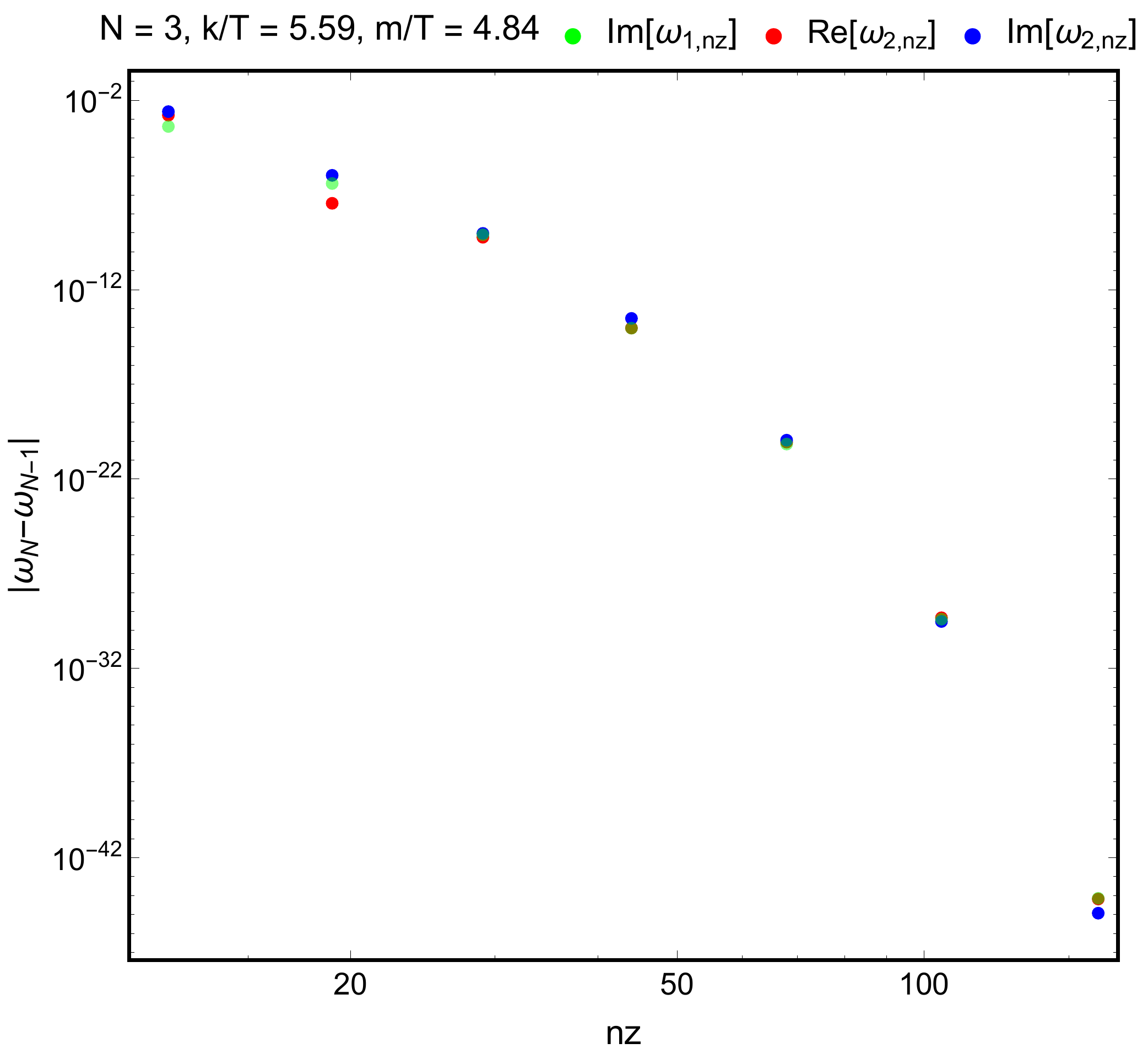}
\quad
\includegraphics[width=6cm]{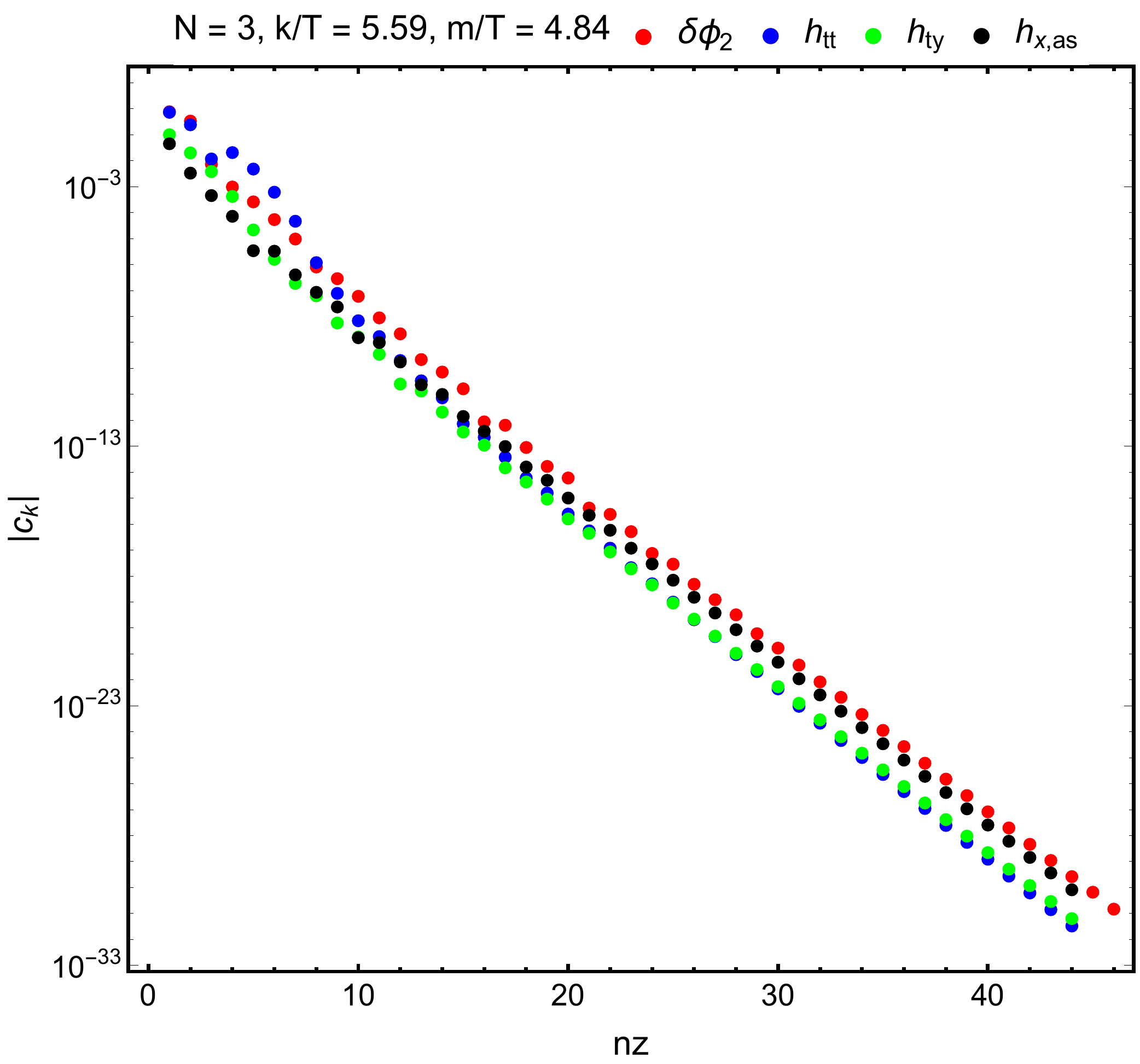}
\caption{\textbf{Left: }Moving of the lowest three QNMs with increasing grid size. \textbf{Right: }Chebychev-coefficients of the eigenfunctions corresponding to the second lowest QNM. \label{pic:conv}}
\end{figure}

We checked that our solutions fulfill the equations of motions and all constraint equations. To check the convergence of the numerical solution, we monitor the change of the solution for finer discretizations. As depicted in the l.h.s. of figure \ref{pic:conv}, the change of the quasi-normal mode frequency decays exponentially with a growing number of grid points; the same is valid for the corresponding eigenfunctions. Another check for the numerical method are the Chebychev-coefficients of the solution, displayed in the r.h.s. of figure \ref{pic:conv}; the coefficients decay exponentially, indicating exponential accuracy of our numerical method.  

\subsection*{Correlation Functions}
In order to calculate the correlation functions, outlined in equation \eqref{corrhydro}, we have to source the scalar field explicitly. Note, that for $k=0$ it is sufficient to consider the $\delta\phi_2-h_{ty}$ sector since the $h_{tt}$ and $h_{x,a}$ decouple.  The corresponding asymptotic expansions are of the form\footnote{We assume that $N>5/2$. For $N < 5/2$ the term proportional to $u^{5-2N}$ corresponds to the expectation value  and the logarithmic term is proportional to the same power in $u$.}
\begin{align}\label{eq:bdyexp}
&\delta\phi_2\sim \phi_\text{vev}+  \phi_\text{log} \log(u)+\ldots+\phi_\text{source} u^{5-2N}\\
&h_{ty}\sim h_{ty,\text{source}}+ \ldots+h_{ty,\text{vev}} u^{3},
\end{align}
where we set $h_{ty,\text{source}}=0$ and $\phi_\text{source}=1$. Note, that the logarithmic terms are not present for $N=3$ and $N \notin \mathbb Z/2$. In order to ensure convergence in presence of logarithmic term (e.g. in the cases $N=4,5,6,...$), we use a suitable mapping for the radial coordinate $u$, for instance $u\mapsto u^2$ (for more details see \cite{Grieninger:2017jxz,Ammon:2016fru}).

\subsection*{Green's Function}
The retarded Green's function  of operators $\mathcal{O}_a$ and $\mathcal{O}_b$ is given by
\begin{equation}\label{eq:greendef}
  \mathcal{G}_{\mathcal{O}_a \mathcal{O}_b}(t-t', \vec{x}-\vec{x}') = -i \theta(t-t') \left\langle [\mathcal{O}_a(t,\vec{x}), \mathcal{O}_b(t',\vec{x}')] \right\rangle \, .
\end{equation}
The Fourier-transformed retarded Green's function $\tilde{\mathcal{G}}_{\mathcal{O}_a \mathcal{O}_b}(\omega, \vec{k})$ reads\footnote{From now on, we drop the tilde to simplify notation.}
\begin{equation}
    \tilde{\mathcal{G}}_{\mathcal{O}_a \mathcal{O}_b}(\omega, \vec{k}) = \int \mathrm{d}t \, \mathrm{d}^d{x} \, e^{-i \omega t + i\vec{k}\cdot\vec{x}} \mathcal{G}_{\mathcal{O}_a \mathcal{O}_b}(t, \vec{x}) \, ,
\end{equation}
and it describes the linear response to $ \delta\langle \mathcal{O}_a(\omega,\vec{k})\rangle$ for a given perturbation $j_b(\omega,\vec{k})$ by\footnote{The coupling of $j_b$ to $\mathcal{O}_b$ is described by replacing the QFT Lagrangian $\mathcal{L}$ by $\mathcal{L} + j_b(x) \mathcal{O}_b(x).$} 
\begin{equation}
    \delta\langle  \mathcal{O}_a(\omega,\vec{k})\rangle = - \mathcal{G}_{\mathcal{O}_a \mathcal{O}_b}(\omega, \vec{k}) \, j_b(\omega,\vec{k}) \, .
\end{equation}
Using holographic techniques the retarded Green's functions are given by
\begin{align}
\mathcal{G}_{T_{\mu\nu} T_{\rho\sigma}}(\omega,\vec{k}) &= - \frac{3}{2}\, \frac{h_{\mu\nu \, \textrm{vev}}}{h^{\rho\sigma}_{\textrm{source}}} \, , \\
\mathcal{G}_{\phi\phi}(\omega,\vec{k}) &=- N \, (2N-5)\, m^2 \, \frac{\phi_{\textrm{vev}}}{\phi_{\textrm{source}}} \, , \\
\mathcal{G}_{T_{\mu\nu} \phi}(\omega,\vec{k}) &=  - \frac{3}{2}\, \frac{h_{\mu\nu \, \textrm{vev}}}{\phi_{\textrm{source}}} \, ,  
\end{align}
where $h_{\mu\nu \, \textrm{vev}}$, $h^{\rho\sigma}_{\textrm{source}}$, $\phi_{\textrm{vev}}$ and $\phi_{\textrm{source}}$ are the boundary values as given in \eqref{eq:bdyexp}. The prefactors of the Green's function $\mathcal{G}_{\phi\phi}$ may be explained as follows: the multiplicative factor $2\Delta -d$ arising in holographic renormalisation reduces to $2N-5$ for the $(2+1)$-dimensional QFT considered here. The factors $N m^2$ are due to the non-canonical normalisation of the kinetic term for the fluctuations of the scalar field. In particular, the prefactor of the kinetic term reads $m^2 \, V'(\bar{X})$ where $\bar{X}$ denotes $X$ evaluated on the background. 

Next we determine the correct normalization of the Goldstone operator. Note that the Goldstone operator $\Phi^I$ has to satisfy $\mathcal{G}_{T_{ty} \Phi^{(2)}}(\omega,\vec{k}=0) = i/\omega$, while on the gravity side we numerically confirmed
\begin{equation}
    \mathcal{G}_{T_{ty} \phi^{(2)}}(\omega, \vec{k}=0) = - \frac{i}{\omega} \, N (2N-5) m^2 \, .
\end{equation}
Hence we can establish the following map between the Goldstone operator $\Phi^I$, on the quantum field theory side, and the St\"uckelberg field $\phi^I$ on the gravity side,
\begin{equation}
    \phi^I = - N (2N-5) \, m^2  \,\Phi^I \, .
\end{equation}

\section{The Hydrodynamic Theory}\label{hydroapp}
In this section we present the hydrodynamic calculations which lead to the dispersion relations used in this paper. We follow the reasoning of \cite{Delacretaz:2017zxd} and expand on their results, focusing on the case of zero density. First we present the effective hydrodynamic theory and then compute the dispersion relations in the transverse and longitudinal sector.

\subsection{Set-up}

We will consider the linearised hydrodynamics of small fluctuations around a $2+1$ dimensional system at equilibrium. Without loss of generality we only allow excitations to depend on $x$ and $t$, hence the momentum is in the $x$-direction. We also adopt the convention of denoting the $x$-component of a quantity by $\|$, and the $y$-component by $\perp$, to indicate the orientation with respect to the momentum. 

The hydrodynamic variables which are of interest for our analysis are
\begin{equation}\label{eq:hydrovariables}
    \varepsilon(t,{x}), \quad  \pi_\|(t,{x}), \quad \pi_\perp(t,{x}), \quad \lambda_\|(t,{x}), \quad \lambda_\perp(t,{x}), 
\end{equation}
for, in order, energy density; momentum density parallel and transverse to the momentum; and $\lambda_\|(t,{x})$, $\lambda_\perp(t,{x})$ are the divergence and curl of the two-component Goldstone field $\Phi(t,{x})$, respectively (not to be confused with the scalar bulk fields $\phi^I$). To be more precise, for the system at hand, the field $\Phi(t,\vec{x})$ is a vector of the form
\begin{equation}
    \Phi(t,\vec{x})= \begin{pmatrix}
    \Phi^{(1)}(t,\vec{x}) \\ \Phi^{(2)}(t,\vec{x})
    \end{pmatrix}.
\end{equation}
From the above Goldstone field we define 
\begin{align}\label{eq:lambdaDEF}
    \lambda_\|(t,\vec{x}) = \partial_x \Phi^{(1)}(t,\vec{x}) + \partial_y \Phi^{(2)}(t,\vec{x}), \quad    \lambda_\perp(t,\vec{x}) = \partial_x \Phi^{(2)}(t,\vec{x}) - \partial_y \Phi^{(1)}(t,\vec{x}) ,
\end{align}
which reduces to $\lambda_\|(t,{x}) = \partial_x \Phi^{(1)}(t,{x})$ and $\lambda_\|(t,{x}) = \partial_x \Phi^{(2)}(t,{x})$ in the case of $y$-independent fluctuations.

The sources of the theory are 
\begin{equation}\label{eq:hydrosources}
     T(t,x), \quad v_\|(t,{x}), \quad v_\perp(t,{x}),  \quad s_\|(t,{x}), \quad s_\perp(t,{x}),
\end{equation}
which denote the temperature; the parallel and transverse components of $\vec{v}(t,x)$; and linear displacements parallel and transverse to the momentum, respectively. The ordering of the above sources indicates the corresponding variable when compared to the list \eqref{eq:hydrovariables}. 

The hydrodynamic variables, collectively denoted $\varphi_a$, are related to the their corresponding sources, $s^b$, through the thermodynamic susceptibilities $\chi_{ab}$ via the following equation
\begin{equation}\label{eq:suscept}
    \varphi_a = \chi_{ab} s^b, \quad \mathrm{with}\quad  \chi_{ab} = \frac{\partial \varphi_a}{\partial s^b}.
\end{equation}
In the linear hydrodynamic regime the susceptibilities will be between small fluctuations of the variables and sources. Hence, for the system under consideration, the above relations can be expressed in matrix form as follows
\begin{equation}\label{eq:suscebtibilitymatrixFULL}
    \begin{pmatrix}
    \delta \varepsilon(t,{x}) \\ \delta \pi_\|(t,{x}) \\ \delta \pi_\perp(t,{x}) \\ \delta \lambda_\|(t,{x}) \\ \delta \lambda_\perp (t,{x})
    \end{pmatrix}
    =
    \begin{pmatrix}
    
    \chi_{\varepsilon T} & 0 & 0 & 0 & 0 \\
     0 & \chi_{\pi\pi} & 0 & 0 & 0 \\
    0 & 0 & \chi_{\pi\pi} & 0 & 0 \\
    0 & 0 & 0 & \chi_{\lambda_\| s_\|} & 0 \\
    0 & 0 & 0 & 0 & \chi_{\lambda_\perp s_\perp}
    \end{pmatrix}
    \begin{pmatrix}
     \delta T(t,{x}) \\ \delta v_\|(t,{x}) \\ \delta v_\perp(t,{x}) \\ \delta s_\|(t,{x}) \\  \delta s_\perp(t,{x})
    \end{pmatrix},
\end{equation}
where we have assumed that the susceptibility between momentum and velocity is identical in the parallel and transverse sectors. In the above equation it is manifest that the translation from sources to variables is facilitated by the inverse of the expressed susceptibility matrix.

Conservation of energy and momentum provides us with the equations
\begin{align}
    \dot{\varepsilon}(t,{x}) + \partial^i \tensor{\tau}{^0_i}(t,{x}) &= 0, \label{eq:entropyconservation} \\
    \dot{\pi}_i(t,{x}) + \partial^j \tau_{ij}(t,{x}) &= 0, \label{eq:momentumconservation}
\end{align}
where a dot above a variable indicates time-derivative. The constitutive relation for $\tau_{i0}(t,{x})$ reads, to first order in derivatives, 
\begin{equation}\label{eq:heatcurrent}
    \tensor{\tau}{^0_i}(t,{x}) = \chi_{\pi\pi} v_i(t,{x}) - \Bar{\kappa}_0 \partial_i T(t,{x}) - T(t,x) \gamma_2 \partial_i s_\|(t,{x}) + \mathcal{O}(\partial^2) ,
\end{equation}
where $\Bar{\kappa}_0$ and $\gamma_2$ are transport coefficients.\footnote{Note that the given expression for the time components of the stress-tensor is equivalent to the heat current at zero charge density, given in \cite{Delacretaz:2017zxd}.} Furthermore, the spatial component of the stress-tensor, $\tau_{ij}(t,x)$, to first order of derivatives, is expressed by
\begin{equation}\label{eq:stresstensorFULL}
\begin{split}
      \tau_{ij}(t,{x}) &= \delta_{ij}\left[ p(t,{x}) - (\kappa + G) \partial\cdot\phi(t,{x}) \right] \\ &\quad -2G\left[ \partial_{(i}\phi_{j)}(t,{x}) - \delta_{ij} \partial\cdot\phi(t,{x}) \right]- \sigma_{ij}(t,{x}) + \mathcal{O}(\partial^2),  
\end{split}
\end{equation}
where $p(t,{x})$ is the thermodynamic pressure; $\kappa$ and $G$ are the bulk and shear elastic modulus, respectively; and 
\begin{equation}\label{eq:sigmaij}
    \sigma_{ij}(t,{x}) = \eta \del{\partial_i v_j(t,{x}) + \partial_j v_i(t,{x}) - \delta_{ij} \partial_k v^k(t,{x})},
\end{equation}
is the dissipative term, with $\eta$ being the shear viscosity.

In addition to the conservation equations there are also two Josephson relations. For $\lambda_\perp(t,x)$ the Josephson relation is given by
\begin{equation}\label{eq:josephsonPerp}
    \dot{\lambda}_\perp(t,{x}) - \partial\times \vec{v}(t,{x}) - \xi_\perp \partial_i\partial^i s_\perp(t,{x}) = 0;
\end{equation}
and for $\lambda_\|(t,x)$ the relation is
\begin{align}\label{eq:josephsonPara}
    \dot{\lambda}_\|(t,{x}) - \partial\cdot \vec{v}(t,{x}) - \gamma_2  \partial_j\partial^j T(t,{x}) - \xi_\|  \partial_k\partial^k s_\|(t,{x}) = 0,
\end{align}
where $\xi_\perp$ and $\xi_\|$ are transport coefficients.\footnote{In the main text we do not distinguish between $\xi_\|$ and $\xi_\perp$, instead we denote the relevant parameter by $\xi$.}


\subsection{Dispersion Relations of Transverse Sector}\label{sec:TransverseDispersion}

Although the main analysis of this paper is concerned with the dispersion relations of the longitudinal sector of the hydrodynamics, it is nevertheless beneficial to establish the techniques used to compute the quantities under investigation by first considering the simpler case of the transverse sector. 

In the transverse sector the hydrodynamic variables are $\pi_\perp(t,{x})$ and $\lambda_\perp(t,{x})$, with corresponding sources $v_\perp(t,{x})$ and $s_\perp(t,{x})$, respectively. To linear order in fluctuations the expansions are 
\begin{align}
    \pi_\perp (t,{x}) &= \Bar{\pi}_\perp + \delta \pi_\perp(t,{x}) + \mathcal{O}(\delta^2), \\
    \lambda_\perp (t,{x}) &= \Bar{\lambda}_\perp + \delta \lambda_\perp(t,{x}) + \mathcal{O}(\delta^2) ,
\end{align}
for the variables, and
\begin{align}
    v_\perp (t,{x}) &= \Bar{v}_\perp + \delta v_\perp(t,{x}) + \mathcal{O}(\delta^2), \\
    s_\perp (t,{x}) &= \Bar{s}_\perp + \delta s_\perp(t,{x}) + \mathcal{O}(\delta^2), 
\end{align}
for the sources. The barred quantities represent the constant equilibrium values upon which we are perturbing. In the present sector the susceptibility relations \eqref{eq:suscebtibilitymatrixFULL} reduce to the two independent equations
\begin{equation}\label{eq:susceptmatrixtrans}
    \begin{pmatrix}
    \delta \pi_\perp(t,{x}) \\ \delta \lambda_\perp(t,{x}) 
    \end{pmatrix}
    =
    \begin{pmatrix}
    \chi_{\pi\pi} & 0 \\ 0 & \chi_{\lambda_\perp s_\perp} 
    \end{pmatrix}
    \begin{pmatrix}
    \delta v_\perp(t,{x}) \\ \delta s_\perp(t,{x}) 
    \end{pmatrix},
\end{equation}
with
\begin{align}
    \chi_{\lambda_\perp s_\perp} &= \frac{1}{G},
\end{align}
where $G$ is the shear elastic modulus.


The only conservation equation of relevance for the transverse sector is that of momentum conservation \eqref{eq:momentumconservation}, which under the circumstances for the linearised system reads
\begin{equation}
    \delta\dot{\pi}_\perp(t,{x}) + \partial_x \tau_{xy}(t,{x}) = 0,
\end{equation}
where, evidently, the only spatial derivative that contributes is with respect to $x$. Using the constitutive relation of the stress-tensor \eqref{eq:stresstensorFULL} we get
\begin{equation}\label{eq:symstresstensorsource}
    \tau_{xy}(t,x) = -G\lambda_\perp(t,x) - \sigma_{xy}(t,x) + \mathcal{O}(\delta^2),
\end{equation}
where we have used $\partial_x \Phi^{(2)}(t,x)= \lambda_\perp(t,x)$; and the dissipative term $\sigma_{xy}(t,x)$ is given by equation \eqref{eq:sigmaij}, which yields
\begin{equation}
    \sigma_{xy}(t,x) = \eta \partial_x v_\perp (t,x).
\end{equation}
At linear order the derivative of the stress-tensor is equal to
\begin{equation}
    \partial_x \tau_{xy}(t,x) = -G \partial_x\delta \lambda_\perp(t,x) - \eta \partial^2_x \delta v_\perp(t,x) + \mathcal{O}(\delta^2).
\end{equation}
Using the inverse of \eqref{eq:susceptmatrixtrans} the velocity fluctuation can be related to the momentum density, hence we arrive at
\begin{equation}
    \partial_x \tau_{xy}(t,x) = -G \partial_x\delta \lambda_\perp(t,x) - \frac{\eta}{\chi_{\pi\pi}} \partial^2_x \delta \pi_\perp(t,x) + \mathcal{O}(\delta^2).
\end{equation}
The momentum conservation equation, up to linear order in fluctuations of the hydrodynamic variables, thus reads
\begin{equation}\label{eq:linearmomentumconsPERP}
    \partial_t \delta {\pi}_\perp(t,x) -G \partial_x\delta \lambda_\perp(t,x) - \frac{\eta}{\chi_{\pi\pi}} \partial^2_x \delta \pi_\perp(t,x) = 0.
\end{equation}

In addition to the conservation equation we need to consider the Josephson relation \eqref{eq:josephsonPerp}. In the same way as for the stress-tensor, linearising the equation and expressing it in terms of the thermodynamic variables, yields
\begin{equation}\label{eq:linearjosephsonPERP}
    \partial_t \delta {\lambda}_\perp(t,x) - \frac{1}{\chi_{\pi\pi}} \partial_x \delta \pi_\perp (t,x) - G \xi_\perp \partial^2_x \delta \lambda_\perp (t,x) = 0.
\end{equation}



The first step in the pursuit of the dispersion relations is to evaluate the derivatives in the equation of momentum conservation and the Josephson relation. This requires that we Fourier transform and identify coefficients. Doing so, we arrive at the system of equations 
\begin{align}
    \del{-i\omega + \frac{\eta}{\chi_{\pi\pi}}k^2  }\delta \pi_\perp(\omega,k) - i G  k \delta \lambda_\perp (\omega, k)  &= 0, \\
    \del{-i \omega + G \xi_\perp k^2} \delta \lambda_\perp (\omega, k) - \frac{ik}{\chi_{\pi\pi}}\delta \pi_\perp (\omega, k) & = 0,
\end{align}
where $\omega$ is the frequency and $k$ is the momentum in the $x$-direction. The above system can be written in matrix form as
\begin{equation}
    \begin{pmatrix}
    -i\omega + {\eta k^2}/{\chi_{\pi\pi}} & - {i G k} \\
    - {ik}/{\chi_{\pi\pi}} & -i \omega + {G \xi_\perp k^2}
    \end{pmatrix}
    \begin{pmatrix}
    \delta \pi_\perp(\omega,k) \\ \delta \lambda_\perp (\omega, k)
    \end{pmatrix}
    = 
     \begin{pmatrix}
    0 \\ 0
    \end{pmatrix},
\end{equation}
which has non-trivial solutions only if the matrix on the left-hand side is non-invertible, i.e. if its determinant is zero. Setting this constraint means
\begin{equation}
 \omega^2 + i \omega \del{G \xi_\perp + \frac{\eta}{\chi_{\pi\pi}} }k^2  - \del{ \frac{G}{\chi_{\pi\pi}}+\frac{\eta G \xi_\perp}{\chi_{\pi\pi} }k^2 }k^2 \overset{!}{=} 0 .
\end{equation}
Solving the above equation for $\omega$, taking the limit of small momentum, and comparing to the known form of a propagating mode with diffusion
\begin{equation}
    \omega = \pm c_T k - i D_T k^2,
\end{equation}
we identify the speed of sound of the transverse mode
\begin{equation}\label{eq:speedperp}
    c_T^2 =  \frac{G}{\chi_{\pi\pi} } ,
\end{equation}
and diffusion constant 
\begin{equation}
    D_T = \frac{1}{2}\del{{G \xi_\perp} + \frac{\eta}{\chi_{\pi\pi}} }.
\end{equation}
This concludes the calculation of the dispersion relations in the transverse sector.


\subsection{Dispersion Relations of Longitudinal Sector}
The procedure of finding the dispersion relations in the longitudinal sector is in principle the same as for the transverse sector. The main differences stem  from a larger set of hydrodynamic variables, which in turn provide us with more conservation equations and thus more hydrodynamic modes. 

The variables for the parallel sector are: energy density $\varepsilon(t,{x})$; momentum $\pi_\|(t,{x})$; and field component $\lambda_\|(t,{x})$. Respectively, the sources are the temperature $T(t,{x})$; velocity $v_\|(t,{x})$; and displacement $s_\|(t, {x})$. 
For the given set of variables and sources the conservation equations are
\begin{align}
    \dot{\varepsilon}(t,{x}) + \partial_x \tensor{\tau}{^0_x}(t,{x}) &= 0, \label{eq:energyconsvpara}\\
    \dot{\pi}_\|(t,{x}) + \partial_x \tau_{xx}(t,{x}) &= 0, \label{eq:momentumconsvpara}
\end{align}
with the addition of the Josephson relation \eqref{eq:josephsonPara}. The linearised variables take the form
\begin{align}
    \varepsilon(t,{x}) &= \bar{\varepsilon} + \delta \varepsilon(t,x) + \mathcal{O}(\delta^2), \\
    \pi_\|(t,{x}) &= \bar{\pi}_\| + \delta \pi_\|(t,x) + \mathcal{O}(\delta^2) ,\\
    \lambda_\|(t,{x}) &= \bar{\lambda}_\| + \delta \lambda_\|(t,x) + \mathcal{O}(\delta^2).
\end{align}
The corresponding sources become, to first order in fluctuations,
\begin{align}
    T(t,{x}) &= \bar{T} + \delta T(t,x) + \mathcal{O}(\delta^2), \\
    v_\|(t,{x}) &= \bar{v}_\| + \delta v_\|(t,x) + \mathcal{O}(\delta^2) ,\\
    s_\|(t,{x}) &= \bar{s}_\| + \delta s_\|(t,x) + \mathcal{O}(\delta^2).
\end{align}
Furthermore, the susceptibility relations for the fluctuations read
\begin{equation}\label{eq:suscebtibilitymatrixpara}
    \begin{pmatrix}
    \delta \varepsilon(t,{x})  \\ \delta \lambda_\|(t,{x}) \\ \delta \pi_\|(t,{x})
    \end{pmatrix}
    =
    \begin{pmatrix}
     \chi_{\varepsilon T} & 0 & 0   \\
    0 & \chi_{\lambda_\| s_\|} & 0   \\
    0 & 0 & \chi_{\pi\pi}    \\
    \end{pmatrix}
    \begin{pmatrix}
     \delta T(t,{x})  \\ \delta s_\|(t,{x})  \\ \delta v_\|(t,{x})
    \end{pmatrix},
\end{equation}
with values of the coefficients being
\begin{align}
    \chi_{\varepsilon T} = c_V, \quad \chi_{\lambda_\| s_\|} = \frac{1}{\kappa + G},
\end{align}
where $c_V$ denotes the specific heat $c_V = {{\partial \varepsilon}/{\partial T}}$.



We start by considering the equation of energy conservation \eqref{eq:energyconsvpara}. The temporal component of the stress-tensor is given by \eqref{eq:heatcurrent}, hence at linear order we have
\begin{equation}
    \partial_x \tensor{\tau}{^0_x}(t,{x}) = \chi_{\pi\pi} \partial_x \delta v_\|(t,{x}) - \Bar{\kappa}_0 \partial^2_x \delta T(t,{x}) -\gamma_2 \bar{T} \partial^2_x \delta s_\|(t,{x}) + \mathcal{O}(\delta^2).
\end{equation}
Using the susceptibilities to write the stress-tensor in terms of hydrodynamic variables, Fourier transforming, and evaluating derivatives, gives the expression for the energy conservation equation
\begin{equation}\label{eq:energyparafinal}
    \del{-i\omega + \frac{\bar{\kappa}_0}{c_V} k^2 }\delta \varepsilon(\omega,{k}) + ik \delta \pi_\|(\omega,{k}) + \gamma_2 (\kappa + G) \bar{T} k^2 \delta \lambda_\|(\omega,{k}) = 0.
\end{equation}

Moving on, let us turn our attention to the equation of momentum conservation \eqref{eq:momentumconsvpara}. The spatial stress-tensor components are given by equation \eqref{eq:stresstensorFULL}, which at linear order gives
\begin{equation}
   \partial_x \tau_{xx}(t,x) = \partial_x \delta p(t,x) - (\kappa + G) \partial_x \delta \lambda_\| (t,x)- \eta  \partial^2_x v_\|(t,x) + \mathcal{O}(\delta^2),
\end{equation}
where we used $\partial_x \Phi^{(1)}(t,x) = \lambda_\|(t,x)$. In terms of the hydrodynamic variables the above equation is equal to
\begin{equation}
    \partial_x \tau_{xx}(t,x) = \frac{\partial p}{\partial \varepsilon} \partial_x \delta \varepsilon(t,x) - (\kappa + G) \partial_x \delta \lambda_\|(t,x) - \frac{\eta}{\chi_{\pi\pi}}  \partial^2_x \delta \pi_\|(t,x). 
\end{equation}
Fourier-transforming and differentiating the momentum conservation equation thus yields
\begin{equation}\label{eq:momentumparafinal}
    \del{-i\omega + \frac{\eta}{\chi_{\pi\pi}} k^2 }\delta \pi_\|(\omega,k) +v^2_u ik \delta \varepsilon(\omega,k) - ik (\kappa + G)  \delta \lambda_\|(\omega,k) = 0.
\end{equation}

Finally, following all the above steps for re-expressing the Josephson relation, yields at linear order
\begin{equation}\label{eq:josephsonparafinal}
     \del{-i\omega + \xi_\| (\kappa+G) k^2}\delta \lambda_\|(\omega,k) - \frac{ik}{\chi_{\pi \pi }} \delta \pi_\|(\omega,k)  + \frac{\gamma_2}{c_V} k^2  \delta \varepsilon(\omega,k) = 0.
\end{equation}

In the same fashion as for the transverse sector, the equations \eqref{eq:energyparafinal}, \eqref{eq:momentumparafinal} and \eqref{eq:josephsonparafinal} can be expressed as a matrix whose determinant must be zero in order to yield non-trivial solutions. This constraint on the determinant leads to a cubic equation in $\omega$, hence the system we are considering will have three frequency modes. Two propagating modes (moving in opposite directions) and one diffusive mode. The generic dispersion relation of the longitudinal propagating mode is
\begin{equation}
    \omega = \pm c_L k - i D_p  k^2 .
\end{equation}
The speed of sound for this mode is given by
\begin{equation}\label{eq:speedlongi}
    c_L^2 = \frac{\partial p}{ \partial \varepsilon} + \frac{\kappa +G}{\chi_{\pi\pi}}.
\end{equation}
Moreover, the dampening is
\begin{equation}
    D_p = \frac{1}{2}\frac{\eta}{\chi_{\pi\pi}} + \frac{1}{2}\frac{c_V (\kappa+G)^2\xi_\| - (\kappa+G) c_V (\partial p/\partial \varepsilon) \bar{T} \gamma_2 + (\partial p/\partial \varepsilon)  \bar{\kappa}_0 \chi_{\pi\pi} -\gamma_2 (\kappa+G) \chi_{\pi\pi}}{c_V (\kappa + G + (\partial p/\partial \varepsilon)  \chi_{\pi\pi})}.
\end{equation}

The non-progapating mode has the dispersion relation
\begin{equation}
    \omega = -i D_\Phi k^2,
\end{equation}
with diffusion constant given by
\begin{equation}
    D_\Phi = (\kappa + G) \frac{\bar{\kappa}_0 + \gamma_2 \chi_{\pi\pi} + c_V (\partial p/\partial \varepsilon)  (\bar{T} \gamma_2 + \xi_\| \chi_{\pi\pi}) }{c_V (\kappa + G + (\partial p/\partial \varepsilon)  \chi_{\pi\pi})}.
\end{equation}
We analyse the above modes in the main text, within the context of our holographic model.\footnote{In order to simplify notation we, in the main text, replace $\bar{T} \to T$ when the parameter appears in a formula.}  

\subsection{Kubo Formulas}\label{sec:Kuboformulae}

The susceptibilities and conservation equations presented in the above subsection can be used in the standard machinery for obtaining retarded Green's functions $\mathcal{G}^R_{\mathcal{O}_a \mathcal{O}_b}(\omega,k)$. In turn, these Green's functions provide us with the following, especially relevant, Kubo formulas
\begin{align}
    \eta &= -\lim_{\omega \to 0} \omega \lim_{k \to 0} \frac{1}{k^2} \mathrm{Im}[\mathcal{G}^R_{\pi_\perp \pi_\perp}(\omega,k)],\\
    G &= \lim_{\omega \to 0} \omega^2 \lim_{k \to 0} \frac{1}{k^2} \mathrm{Re}[\mathcal{G}^R_{\pi_\perp \pi_\perp}(\omega,k)],\\
    \bar{\kappa}_0 &= -\lim_{\omega \to 0} \omega \lim_{k \to 0} \frac{1}{k^2} \mathrm{Im}[\mathcal{G}^R_{\varepsilon\varepsilon}(\omega,k)],\\
    \xi_\| &= \lim_{\omega \to 0} \omega \lim_{k \to 0}  \mathrm{Im}[\mathcal{G}^R_{\Phi^{(1)} \Phi^{(1)}}(\omega,k)],\\
    \bar{T} \,\gamma_2 &= -\lim_{\omega \to 0} \omega \lim_{k \to 0} \frac{1}{k} \mathrm{Re}[\mathcal{G}^R_{\varepsilon\Phi^{(1)}}(\omega,k)], \\
    \chi_{\pi\pi} &= \lim_{\omega \to 0} \omega \lim_{k \to 0} \frac{1}{k} \mathrm{Re}[\mathcal{G}^R_{\varepsilon\pi_\|}(\omega,k)],\\
   \chi_{\pi\pi}\, c_L^2 &= \lim_{\omega\to0} \omega^2 \lim_{k\to0} \frac{1}{k^2} \textrm{Re} [ \mathcal{G}^R_{\pi_\parallel \pi_\parallel}(\omega,k)]. \label{eq:kubosound} 
\end{align}
In the above formulas we have expressed the Green's functions in terms of $\Phi^{(1)}(t,x)$ by relating the Goldstone field to $\lambda_\|(t,x)$, which is done by Fourier-transforming and evaluating the derivative of \eqref{eq:lambdaDEF}, together with the structure of \eqref{eq:greendef}.\footnote{Similarly, Green's functions containing $\pi_\|$ or $\pi_\perp$ can be re-expressed in terms of the stress-tensor by using the equation of momentum conservation.}  Assuming that the speed of sound $c_L^2$ is given by \eqref{eq:speedlongi}, the Kubo formula \eqref{eq:kubosound} provides us with a definition for $\kappa$ via the expression
\begin{equation}
    \kappa = \lim_{\omega\to0} \omega^2 \lim_{k\to0} \frac{1}{k^2} \textrm{Re} [ \mathcal{G}^R_{\pi_\parallel \pi_\parallel}(\omega,k)] - (\partial p/\partial \varepsilon) \, \chi_{\pi\pi} - G \, .
\end{equation}
Further discussions regarding the results obtained from the above Kubo formulas can be found in the main text.

\bibliographystyle{JHEP}
\bibliography{sound}

\providecommand{\href}[2]{#2}\begingroup\raggedright\begin{thebibliography}{10}

\bibitem{bro}
D.~Forster, \emph{{Hydrodynamic Fluctuations, Broken Symmetry, And Correlation
  Functions}}.
\newblock CRC Press.

\bibitem{Nicolis:2015sra}
A.~Nicolis, R.~Penco, F.~Piazza and R.~Rattazzi, \emph{{Zoology of condensed
  matter: Framids, ordinary stuff, extra-ordinary stuff}},
  \href{http://dx.doi.org/10.1007/JHEP06(2015)155}{\emph{JHEP} {\bf 06} (2015)
  155}, [\href{https://arxiv.org/abs/1501.03845}{{\tt 1501.03845}}].

\bibitem{PhysRevB.22.2514}
A.~Zippelius, B.~I. Halperin and D.~R. Nelson, \emph{Dynamics of
  two-dimensional melting},
  \href{http://dx.doi.org/10.1103/PhysRevB.22.2514}{\emph{Phys. Rev. B} {\bf
  22} (Sep, 1980) 2514--2541}.

\bibitem{PhysRevA.6.2401}
P.~C. Martin, O.~Parodi and P.~S. Pershan, \emph{Unified hydrodynamic theory
  for crystals, liquid crystals, and normal fluids},
  \href{http://dx.doi.org/10.1103/PhysRevA.6.2401}{\emph{Phys. Rev. A} {\bf 6}
  (Dec, 1972) 2401--2420}.

\bibitem{Hartnoll:2016apf}
S.~A. Hartnoll, A.~Lucas and S.~Sachdev, \emph{{Holographic quantum matter}},
  \href{https://arxiv.org/abs/1612.07324}{{\tt 1612.07324}}.

\bibitem{Ammon:2015wua}
M.~Ammon and J.~Erdmenger, \emph{{Gauge/gravity duality}}.
\newblock Cambridge University Press, 2015.

\bibitem{zaanen2015holographic}
J.~Zaanen, Y.~Liu, Y.~Sun and K.~Schalm, \emph{Holographic Duality in Condensed
  Matter Physics}.
\newblock Cambridge University Press, 2015.

\bibitem{Policastro:2002se}
G.~Policastro, D.~T. Son and A.~O. Starinets, \emph{{From AdS / CFT
  correspondence to hydrodynamics}},
  \href{http://dx.doi.org/10.1088/1126-6708/2002/09/043}{\emph{JHEP} {\bf 09}
  (2002) 043}, [\href{https://arxiv.org/abs/hep-th/0205052}{{\tt
  hep-th/0205052}}].

\bibitem{Policastro:2002tn}
G.~Policastro, D.~T. Son and A.~O. Starinets, \emph{{From AdS / CFT
  correspondence to hydrodynamics. 2. Sound waves}},
  \href{http://dx.doi.org/10.1088/1126-6708/2002/12/054}{\emph{JHEP} {\bf 12}
  (2002) 054}, [\href{https://arxiv.org/abs/hep-th/0210220}{{\tt
  hep-th/0210220}}].

\bibitem{Leutwyler:1996er}
H.~Leutwyler, \emph{{Phonons as goldstone bosons}}, {\emph{Helv. Phys. Acta}
  {\bf 70} (1997) 275--286}, [\href{https://arxiv.org/abs/hep-ph/9609466}{{\tt
  hep-ph/9609466}}].

\bibitem{Lubensky}
P.~M. Chaikin and T.~C. Lubensky, \emph{Principles of Condensed Matter
  Physics}.
\newblock Cambridge University Press, 1995,
  \href{http://dx.doi.org/10.1017/CBO9780511813467}{10.1017/CBO9780511813467}.

\bibitem{landau7}
L.~D. Landau and E.~M. Lifshitz, \emph{Course of Theoretical Physics, Vol.
  7,Theory of Elasticity}.
\newblock Pergamon Press, 1970.

\bibitem{doi:10.1021/j150454a025}
F.~H. MacDougall, \emph{Kinetic theory of liquids. by j. frenkel.},
  \href{http://dx.doi.org/10.1021/j150454a025}{\emph{The Journal of Physical
  and Colloid Chemistry} {\bf 51} (1947) 1032--1033},
  [\href{https://arxiv.org/abs/https://doi.org/10.1021/j150454a025}{{\tt
  https://doi.org/10.1021/j150454a025}}].

\bibitem{noirez2012identification}
L.~Noirez and P.~Baroni, \emph{Identification of a low-frequency elastic
  behaviour in liquid water}, {\emph{Journal of Physics: Condensed Matter} {\bf
  24} (2012) 372101}.

\bibitem{PhysRevLett.118.215502}
C.~Yang, M.~T. Dove, V.~V. Brazhkin and K.~Trachenko, \emph{Emergence and
  evolution of the $k$ gap in spectra of liquid and supercritical states},
  \href{http://dx.doi.org/10.1103/PhysRevLett.118.215502}{\emph{Phys. Rev.
  Lett.} {\bf 118} (May, 2017) 215502}.

\bibitem{2016RPPh...79a6502T}
K.~{Trachenko} and V.~V. {Brazhkin}, \emph{{Collective modes and thermodynamics
  of the liquid state}},
  \href{http://dx.doi.org/10.1088/0034-4885/79/1/016502}{\emph{Reports on
  Progress in Physics} {\bf 79} (Jan., 2016) 016502},
  [\href{https://arxiv.org/abs/1512.06592}{{\tt 1512.06592}}].

\bibitem{Baggioli:2018vfc}
M.~Baggioli and K.~Trachenko, \emph{{Solidity of liquids: How Holography knows
  it}},  \href{https://arxiv.org/abs/1807.10530}{{\tt 1807.10530}}.

\bibitem{Baggioli:2018nnp}
M.~Baggioli and K.~Trachenko, \emph{{Maxwell interpolation and close
  similarities between liquids and holographic models}},
  \href{https://arxiv.org/abs/1808.05391}{{\tt 1808.05391}}.

\bibitem{Kovtun:2012rj}
P.~Kovtun, \emph{{Lectures on hydrodynamic fluctuations in relativistic
  theories}}, \href{http://dx.doi.org/10.1088/1751-8113/45/47/473001}{\emph{J.
  Phys.} {\bf A45} (2012) 473001}, [\href{https://arxiv.org/abs/1205.5040}{{\tt
  1205.5040}}].

\bibitem{Delacretaz:2017zxd}
L.~V. Delacrétaz, B.~Goutéraux, S.~A. Hartnoll and A.~Karlsson, \emph{{Theory
  of hydrodynamic transport in fluctuating electronic charge density wave
  states}}, \href{http://dx.doi.org/10.1103/PhysRevB.96.195128}{\emph{Phys.
  Rev.} {\bf B96} (2017) 195128}, [\href{https://arxiv.org/abs/1702.05104}{{\tt
  1702.05104}}].

\bibitem{Davison:2016hno}
R.~A. Davison, L.~V. Delacrétaz, B.~Goutéraux and S.~A. Hartnoll,
  \emph{{Hydrodynamic theory of quantum fluctuating superconductivity}},
  \href{http://dx.doi.org/10.1103/PhysRevB.96.059902,
  10.1103/PhysRevB.94.054502}{\emph{Phys. Rev.} {\bf B94} (2016) 054502},
  [\href{https://arxiv.org/abs/1602.08171}{{\tt 1602.08171}}].

\bibitem{Alberte:2017oqx}
L.~Alberte, M.~Ammon, M.~Baggioli, A.~Jiménez-Alba and O.~Pujolàs,
  \emph{{Holographic Phonons}},  \href{https://arxiv.org/abs/1711.03100}{{\tt
  1711.03100}}.

\bibitem{Baggioli:2014roa}
M.~Baggioli and O.~Pujolas, \emph{{Electron-Phonon Interactions,
  Metal-Insulator Transitions, and Holographic Massive Gravity}},
  \href{http://dx.doi.org/10.1103/PhysRevLett.114.251602}{\emph{Phys. Rev.
  Lett.} {\bf 114} (2015) 251602}, [\href{https://arxiv.org/abs/1411.1003}{{\tt
  1411.1003}}].

\bibitem{Alberte:2015isw}
L.~Alberte, M.~Baggioli, A.~Khmelnitsky and O.~Pujolas, \emph{{Solid Holography
  and Massive Gravity}},
  \href{http://dx.doi.org/10.1007/JHEP02(2016)114}{\emph{JHEP} {\bf 02} (2016)
  114}, [\href{https://arxiv.org/abs/1510.09089}{{\tt 1510.09089}}].

\bibitem{Andrade:2017cnc}
T.~Andrade, M.~Baggioli, A.~Krikun and N.~Poovuttikul, \emph{{Pinning of
  longitudinal phonons in holographic spontaneous helices}},
  \href{http://dx.doi.org/10.1007/JHEP02(2018)085}{\emph{JHEP} {\bf 02} (2018)
  085}, [\href{https://arxiv.org/abs/1708.08306}{{\tt 1708.08306}}].

\bibitem{Baggioli:2019aqf}
M.~Baggioli, U.~Gran, A.~J. Alba, M.~Tornsö and T.~Zingg, \emph{{Holographic
  Plasmon Relaxation with and without Broken Translations}},
  \href{https://arxiv.org/abs/1905.00804}{{\tt 1905.00804}}.

\bibitem{Baggioli:2019abx}
M.~Baggioli and S.~Grieninger, \emph{{Zoology of Solid \& Fluid Holography :
  Goldstone Modes and Phase Relaxation}},
  \href{https://arxiv.org/abs/1905.09488}{{\tt 1905.09488}}.

\bibitem{Donos:2019hpp}
A.~Donos, D.~Martin, C.~Pantelidou and V.~Ziogas, \emph{{Incoherent
  hydrodynamics and density waves}},
  \href{https://arxiv.org/abs/1906.03132}{{\tt 1906.03132}}.

\bibitem{Donos:2019txg}
A.~Donos, D.~Martin, C.~Pantelidou and V.~Ziogas, \emph{{Hydrodynamics of
  broken global symmetries in the bulk}},
  \href{https://arxiv.org/abs/1905.00398}{{\tt 1905.00398}}.

\bibitem{Baggioli:2015gsa}
M.~Baggioli and D.~K. Brattan, \emph{{Drag phenomena from holographic massive
  gravity}},
  \href{http://dx.doi.org/10.1088/1361-6382/34/1/015008}{\emph{Class. Quant.
  Grav.} {\bf 34} (2017) 015008}, [\href{https://arxiv.org/abs/1504.07635}{{\tt
  1504.07635}}].

\bibitem{Baggioli:2015zoa}
M.~Baggioli and M.~Goykhman, \emph{{Phases of holographic superconductors with
  broken translational symmetry}},
  \href{http://dx.doi.org/10.1007/JHEP07(2015)035}{\emph{JHEP} {\bf 07} (2015)
  035}, [\href{https://arxiv.org/abs/1504.05561}{{\tt 1504.05561}}].

\bibitem{Baggioli:2015dwa}
M.~Baggioli and M.~Goykhman, \emph{{Under The Dome: Doped holographic
  superconductors with broken translational symmetry}},
  \href{http://dx.doi.org/10.1007/JHEP01(2016)011}{\emph{JHEP} {\bf 01} (2016)
  011}, [\href{https://arxiv.org/abs/1510.06363}{{\tt 1510.06363}}].

\bibitem{Alberte:2016xja}
L.~Alberte, M.~Baggioli and O.~Pujolas, \emph{{Viscosity bound violation in
  holographic solids and the viscoelastic response}},
  \href{http://dx.doi.org/10.1007/JHEP07(2016)074}{\emph{JHEP} {\bf 07} (2016)
  074}, [\href{https://arxiv.org/abs/1601.03384}{{\tt 1601.03384}}].

\bibitem{Alberte:2017cch}
L.~Alberte, M.~Ammon, M.~Baggioli, A.~Jiménez and O.~Pujolàs, \emph{{Black
  hole elasticity and gapped transverse phonons in holography}},
  \href{http://dx.doi.org/10.1007/JHEP01(2018)129}{\emph{JHEP} {\bf 01} (2018)
  129}, [\href{https://arxiv.org/abs/1708.08477}{{\tt 1708.08477}}].

\bibitem{Ammon:2019wci}
M.~Ammon, M.~Baggioli and A.~Jimenez-Alba, \emph{{A Unified Description of
  Translational Symmetry Breaking in Holography}},
  \href{https://arxiv.org/abs/1904.05785}{{\tt 1904.05785}}.

\bibitem{Baggioli:2018qwu}
M.~Baggioli and A.~Zaccone, \emph{{Universal origin of boson peak vibrational
  anomalies in ordered crystals and in amorphous materials}},
  \href{https://arxiv.org/abs/1810.09516}{{\tt 1810.09516}}.

\bibitem{Skenderis:2002wp}
K.~Skenderis, \emph{{Lecture notes on holographic renormalization}},
  \href{http://dx.doi.org/10.1088/0264-9381/19/22/306}{\emph{Class. Quant.
  Grav.} {\bf 19} (2002) 5849--5876},
  [\href{https://arxiv.org/abs/hep-th/0209067}{{\tt hep-th/0209067}}].

\bibitem{Amoretti:2018tzw}
A.~Amoretti, D.~Areán, B.~Goutéraux and D.~Musso, \emph{{A holographic
  strange metal with slowly fluctuating translational order}},
  \href{https://arxiv.org/abs/1812.08118}{{\tt 1812.08118}}.

\bibitem{Amoretti:2019cef}
A.~Amoretti, D.~Areán, B.~Goutéraux and D.~Musso, \emph{{Diffusion and
  universal relaxation of holographic phonons}},
  \href{https://arxiv.org/abs/1904.11445}{{\tt 1904.11445}}.

\bibitem{Baggioli:2018bfa}
M.~Baggioli and A.~Buchel, \emph{{Holographic Viscoelastic Hydrodynamics}},
  \href{https://arxiv.org/abs/1805.06756}{{\tt 1805.06756}}.

\bibitem{oriol}
M.~Baggioli, V.~Cancer-Castillo and O.~Pujolas, \emph{{To appear}}, .

\bibitem{Esposito:2017qpj}
A.~Esposito, S.~Garcia-Saenz, A.~Nicolis and R.~Penco, \emph{{Conformal solids
  and holography}},
  \href{http://dx.doi.org/10.1007/JHEP12(2017)113}{\emph{JHEP} {\bf 12} (2017)
  113}, [\href{https://arxiv.org/abs/1708.09391}{{\tt 1708.09391}}].

\bibitem{Alberte:2018doe}
L.~Alberte, M.~Baggioli, V.~C. Castillo and O.~Pujolas, \emph{{Elasticity
  bounds from Effective Field Theory}},
  \href{https://arxiv.org/abs/1807.07474}{{\tt 1807.07474}}.

\bibitem{Baier:2007ix}
R.~Baier, P.~Romatschke, D.~T. Son, A.~O. Starinets and M.~A. Stephanov,
  \emph{{Relativistic viscous hydrodynamics, conformal invariance, and
  holography}},
  \href{http://dx.doi.org/10.1088/1126-6708/2008/04/100}{\emph{JHEP} {\bf 04}
  (2008) 100}, [\href{https://arxiv.org/abs/0712.2451}{{\tt 0712.2451}}].

\bibitem{Kim:2016hzi}
K.~K. Kim, M.~Park and K.-Y. Kim, \emph{{Ward identity and Homes’ law in a
  holographic superconductor with momentum relaxation}},
  \href{http://dx.doi.org/10.1007/JHEP10(2016)041}{\emph{JHEP} {\bf 10} (2016)
  041}, [\href{https://arxiv.org/abs/1604.06205}{{\tt 1604.06205}}].

\bibitem{Davison:2014lua}
R.~A. Davison and B.~Goutéraux, \emph{{Momentum dissipation and effective
  theories of coherent and incoherent transport}},
  \href{http://dx.doi.org/10.1007/JHEP01(2015)039}{\emph{JHEP} {\bf 01} (2015)
  039}, [\href{https://arxiv.org/abs/1411.1062}{{\tt 1411.1062}}].

\bibitem{Grozdanov:2018ewh}
S.~Grozdanov and N.~Poovuttikul, \emph{{Generalised global symmetries in states
  with dynamical defects: the case of the transverse sound in field theory and
  holography}},  \href{https://arxiv.org/abs/1801.03199}{{\tt 1801.03199}}.

\bibitem{ishii2019glass}
Y.~Ishii, Y.~Ouchi, S.~Kawaguchi, H.~Ishibashi, Y.~Kubota and S.~Mori,
  \emph{Glass-like features of crystalline solids in the quantum critical
  regime}, {\emph{arXiv preprint arXiv:1901.09502} (2019) }.

\bibitem{setty2019glass}
C.~Setty, \emph{Glass-induced enhancement of superconducting $ t\_c $: Pairing
  via dissipative mediators}, {\emph{arXiv preprint arXiv:1902.00516} (2019) }.

\bibitem{He62}
Y.~He, M.~Hashimoto, D.~Song, S.-D. Chen, J.~He, I.~M. Vishik et~al.,
  \emph{Rapid change of superconductivity and electron-phonon coupling through
  critical doping in bi-2212},
  \href{http://dx.doi.org/10.1126/science.aar3394}{\emph{Science} {\bf 362}
  (2018) 62--65},
  [\href{https://arxiv.org/abs/https://science.sciencemag.org/content/362/6410/62.full.pdf}{{\tt
  https://science.sciencemag.org/content/362/6410/62.full.pdf}}].

\bibitem{Grieninger:2017jxz}
S.~Grieninger, \emph{{Holographic quenches and anomalous transport}},
  \href{https://arxiv.org/abs/1711.08422}{{\tt 1711.08422}}.

\bibitem{Ammon:2016fru}
M.~Ammon, S.~Grieninger, A.~Jimenez-Alba, R.~P. Macedo and L.~Melgar,
  \emph{{Holographic quenches and anomalous transport}},
  \href{http://dx.doi.org/10.1007/JHEP09(2016)131}{\emph{JHEP} {\bf 09} (2016)
  131}, [\href{https://arxiv.org/abs/1607.06817}{{\tt 1607.06817}}].

\end{thebibliography}\endgroup
\end{document}